\documentclass[a4paper,fleqn]{cas-sc}

\usepackage[numbers]{natbib}

\usepackage{amssymb,amsmath,amsfonts,mathrsfs}
\usepackage{xspace,colortbl}
\usepackage{subfigure}
\usepackage{balance}
\usepackage{amssymb}
\usepackage{amsmath}
\usepackage{booktabs}
\usepackage{mathrsfs}
\usepackage{float}
\usepackage{hyperref}
\usepackage{cleveref}
\usepackage[linesnumbered,ruled,vlined]{algorithm2e}

\hypersetup{hypertex=true,
colorlinks=true,
linkcolor=blue,
anchorcolor=blue,
citecolor=blue}
\usepackage{caption}
\makeatletter
\renewcommand{\fnum@figure}{Fig.~{\thefigure}}
\makeatother

\newcommand{\be}{\begin{equation}}
	\newcommand{\ee}{\end{equation}}
\newcommand{\bea}{\begin{eqnarray}}
	\newcommand{\eea}{\end{eqnarray}}

\allowdisplaybreaks[4]

\def\tsc#1{\csdef{#1}{\textsc{\lowercase{#1}}\xspace}}
\tsc{WGM}
\tsc{QE}

\begin{document}
\let\WriteBookmarks\relax
\def\floatpagepagefraction{1}
\def\textpagefraction{.001}
\let\printorcid\relax 

\shorttitle{High-order Finite-Volume Central Targeted ENO Family Scheme for Compressible Flows in Unstructured Meshes}   

\shortauthors{Qihang Ma et al.}

\title[mode = title]{High-order Finite-Volume Central Targeted ENO Family Scheme for Compressible Flows in Unstructured Meshes}  

\tnotemark[1,2]

\author[1]{Qihang Ma}
\author[1,3]{Kai Leong Chong}
\author[2]{Feng Feng}
\author[4]{Jianhua Zhang}
\author[1,5]{Bofu Wang\corref{cor1}}
\author[1,5]{Quan Zhou}
\cortext[cor1]{bofuwang@shu.edu.cn} 



\address[1]{Shanghai Key Laboratory of Mechanics in Energy Engineering, Shanghai Institute of Applied Mathematics and Mechanics, School of Mechanics and Engineering Science, Shanghai University, Shanghai 200072, China}
\address[2]{China Academy of Aerospace Science and Innovation, Beijing 100176, China}
\address[3]{Shanghai Institute of Aircraft Mechanics and Control, Zhangwu Road, Shanghai 200092, China}
\address[4]{School of Microelectronics, Key Laboratory of Advanced Display and System Application of Ministry of Education, \\Shanghai University, Shanghai {\rm200072}, China}
\address[5]{Shanghai Frontier Science Center of Mechanoinformatics , Shanghai University, Shanghai 200072, China}
\cortext[1]{bofuwang@shu.edu.cn qzhou@shu.edu.cn} 

\begin{abstract}
The high-order Target ENO (TENO) scheme, known for its innovative weighting strategy, has demonstrated strong potential for complex flow predictions. This study extends the TENO weighting approach to develop non-oscillatory central TENO (CTENO and CTENOZ) family schemes for unstructured meshes. The CTENO schemes employ compact directional stencils, which increase the likelihood of finding stencils within smooth regions. The design is intentionally compact to simplify the implementation of directional stencils. An effective scale separation strategy is adopted using an ENO-like stencil selection method, which employs large central stencils in smooth regions to achieve high-order accuracy, and smaller directional stencils near discontinuities to improve shock-capturing capabilities. Extensive tests involving CWENO, TENO, CTENO, and CTENOZ schemes were conducted to assess their performance in terms of accuracy, robustness, parallel scalability, and computational efficiency. The findings indicate that the proposed CTENO and CTENOZ schemes deliver high-order precision, lower numerical dissipation, and excellent shock-capturing performance.

\end{abstract}




\maketitle

\section{Introdution}
\label{}

In computational fluid dynamics, high-order accuracy techniques are an essential area of study. It is necessary for the solution to resolve smooth flows with high order precision while keeping the solution non-oscillatory in the presence of discontinuities. High order and low dissipation characteristics are crucial for resolving small-scale flow patterns in complicated flow simulations. These conditions provide challenges for the development of current algorithms for computation. Several methods have been put out for this target, such as artificial viscosity schemes \cite{1}, total variation diminishing (TVD) schemes \cite{2}, essentially non-oscillatory (ENO) schemes \cite{3} and weighted essentially non-oscillatory (WENO) schemes \cite{4}.

WENO schemes, considered as most efficient high-order numerical methods for shock-capturing, were initially proposed in finite-volume format by Liu et al. \cite{4} and subsequently expanded to finite-difference format by Jiang and Shu et al. \cite{5}. These schemes were derived from ENO proposed by Harten et al. \cite{3} and surpass ENO by incorporating all candidate stencils using indicators of smoothness. This guarantees restoration of globally achieving optimal high-order accuracy in smooth areas asymptotically. Subsequently, WENO scheme near critical points was improved by redefining the nonlinear weights, leading to low dissipation such as WENO-M \cite{6} and WENO-Z \cite{7}. However, this practical use of WENO family schemes in accurately predicting complex flows has shown that they introduce excessive dissipation when used for under-resolved simulations, such as turbulence \cite {10}. To enhance the low dissipation characteristic of WENO schemes, several approaches have been explored. These include increasing the non-smooth weights of stencil points \cite {11}, limiting nonlinear adjustments in smooth areas \cite {12}, optimizing spectral characteristics of linear scheme background \cite {13}, and new nonlinear weights for improving accuracy and resolution \cite{58}. For more comprehensive examination of WENO and their respective details, readers are advised to refer \cite {14}.

Fu et al. \cite{15,16} recently explored a group of schemes, known as targeted ENO (TENO). Utilizing ENO-like stencil selection methods, TENO possess low dissipation while still preserving the ability to capture sharp shocks. A notable feature of TENO schemes is their capability to maintain the exact linear scheme for intermediate wavenumbers, without compromising their ability to capture strong discontinuities. TENO schemes on structured meshes have found applications in various flow scenarios \cite{52}, such as multi-phase flows \cite{17}, detonation simulations \cite{18}, and turbulent flows \cite{19,20}. While TENO schemes have been successful in demonstrating their effectiveness for compressible fluid and predicting turbulence on structured grids, extending their applicability to unstructured meshes presents a significant challenge. Simultaneously achieving accuracy with high-order in smooth areas, small structure features with lower dissipation and precise shock-capturing abilities, ensuring numerical stability further compounds the difficulty.

When it comes to handling intricate geometries, it becomes crucial to extend high order schemes to unstructured meshes \cite{59}. As evidenced in \cite{21,22}, when utilizing the traditional WENO method on unstructured grids, like triangular or tetrahedral, optimal linear coefficients needed to create reconstructions with orders by collecting small stencil elements can alter considerably based on different quadrature points and mesh configurations. In certain scenarios, these weights may result in numerical instabilities and necessitate additional specific handling \cite{23}. One possibility is to utilize multi-resolution WENO schemes \cite{25} or adaptive order WENO schemes \cite{26}. One potential technique to alleviate the hardship of defining the ideal linear weights is to lower desired order of accuracy for large stencils to match small candidate stencils \cite{24}. The central WENO (CWENO) schemes \cite{27,54}, incorporate both large and small stencils, implementing customized weighting strategies that permit flexibility in choosing the linear weights. Utilization of the CWENO scheme provides benefits of decreased computational footprints compared with conventional WENO because of smaller directional stencil size. Notably, Tsoutsanis et al. \cite{8} succeed in expanding high-order WENO to unstructured meshes with mixed-element and accurately forecasted complex turbulent flows \cite{9} based on the strategy of weight. Recently, Ji et al. \cite{28,29} extended TENO to unstructured grids within the finite-volume methods, which offers to seventh order accuracy.

Unlike WENO for unstructured grids \cite{8}, CWENO schemes have received less attention despite they have increased robustness. The CWENO reduce width of directional stencils, resulting in more likelihood of one stencil existing in smooth region. This enhanced robustness sets it apart from WENO schemes \cite{54}. In this research, our aim is to develop central target ENO (CTENO) and CTENOZ scheme on unstructured grids under finite volume framework with high-order. Compact design of CTENO scheme makes it particularly well-suited for complex flow simulation. In order to achieve reduced dissipation and improved shock-capture abilities, we utilize stencil selection strategy inspired by ENO method \cite{28}, which is built upon a customized arrangement of candidate stencils, comprising of larger and multiple smaller stencils. The CWENO and CTENO schemes have incorporated central stencil algorithm with stencil-based compact (SBC) and directional stencil Type 3 algorithm \cite{30}, resulting a significant reduction in their computational footprint. The construction of directional stencils included in central stencil is relatively simple and compact (especially for higher-order cases). This simplifies the process compared with original WENO. We have developed the CTENO scheme with a spatial accuracy of up to 7th order for unstructured meshes. Following the WENOZ and CWENOZ scheme \cite{7,8}, the CTENOZ scheme is provided to decrease the numerical dissipation based on CTENO scheme. Our ultimate aim focus on enhancing affordability and robustness of this class of schemes, even in large-scale industrial applications.

This proposed CTENO and CTENOZ scheme is implemented in UCNS3D open source solver \cite{31}. Our main focus is to assess its performance in  accuracy, robustness, parallel scalability and run time for various test benchmarks and apply these methods to more complex flows \cite{60,61,62,63,64,65}. Additionally, we compare the numerical results obtained from CTENO and CTENOZ schemes with TENO and CWENO schemes. The paper is structured as follows: Section 2 introduces numerical framework used to finite-volume methodology with high-order on unstructured grids, including process of reconstruction for WENO, CWENO, TENO, CTENO and CTENOZ, as well as the selected fluxes and temporal discretization methods. In Section 3, we present the results obtained from test benchmarks and compare them with analytical, reference, or experimental solutions. Finally, Section 4 concludes research.

\section{Numerical framework}
\label{}
 
In this study, we examine the Euler equation, which is commonly known as hyperbolic conservation laws. The conservative form of these equations is expressed by
 \begin {equation}
\frac{\partial U}{\partial t} +\bigtriangledown\cdot (F(U))=0,
\label{e1}
\end {equation}
where \(U = U(x,y,z,t)\) is vector of conserved variables,  \(F(U) =( F^{x}(U), F^{y}(U) ,F^{z}(U))\) is convection flux functions. The physical domain comprises a mix of hexahedral,tetrahedral, prism, or pyramid elements in 3D, as well as triangular or quadrilateral elements in 2D. Integrating \autoref{e1} in considered element \(i\) presents in the following format.
\begin {equation}
\frac{\mathrm{d} U_{i} }{\mathrm{d} t}_{}^{}=-\frac{1}{\mid V_{ i}\mid  }     \sum_{j}^{N_{f} } \sum_{k}^{N_{q} }F^{n_{i,j} }(U_{i,j,L}^{n}(x_{i,j,k},t), U_{i,j,R}^{n}(x_{i,j,k},t))\omega _{k} \mid A_{i,j} \mid,  
\end {equation}
where $U_{i}=\frac{1}{\mid V_{i}\mid  }  \int_{V_{i} }^{} U(x,y,z)dV$ is the cell average value of conserved variables, \(V_{i}\) is volume of considered element \(i\), and \(F^{n_{i,j} }\) is numerical flux function in normal direction of interface between cell \(i\) and its adjacent cell \(j\). \(N_f\) is total number of cell faces, \(N_q\) are quadrature points utilized to approximate surface integrals of high-order. \(U_{i,j,L}^{n}\) and \(U_{i,j,R}^{n}\) are the approximate solutions on the interface from the left and right, respectively. To approximate volume/surface/line integrals, Gauss quadrature rule is employed, which suits the polynomial order effectively. \(\omega _{k}\) is weight assigned to Gauss integration point \(x_{i,j,k}\), and \(\mid A_{i,j}\mid\) is surface area of face \(j\). Temporal discretization method employed is the 3rd-order  or 4th-order explicit Runge-Kutta method \cite{38}.

\subsection{Reconstruction}
\label{}

For considered element \(i\), \(r\) order polynomial \(p_{i}(x, y, z)\) can provides \(r + 1 \) order of accuracy, by maintaining that possess an equivalent mean value with \(U_{i}\), which is expressed as:
\begin {equation}
U_{i}=\frac{1}{\mid V_{i}\mid  }  \int_{V_{i} }^{} U(x,y,z)dV=\frac{1}{\mid V_{i}\mid  }  \int_{V_{i} }^{} P_{i}(x,y,z)dV.
\label{e4}
\end {equation}

 For unstructured grids comprising cells of diverse form, converting cells from the physical space \((x, y,z)\) to a reference space  \((\xi ,\eta, \zeta )\) to mitigate scaling effects is advantageous, while ensuring the spatial cell average of conserved variable remains unchanged during the transformation.
 
 \begin {equation}
U_{i}=\frac{1}{\mid V_{i}\mid  }  \int_{V_{i} }^{} U(x,y,z)dV=\frac{1}{\mid V_{i}^{'}  \mid  } \int_{V_{i}^{'}}^{} U(\xi ,\eta ,\zeta )d\xi d\eta d\zeta.
\end {equation}

To compute reconstructed values of specified cell \(i\), compact stencil \(S\) consisting of \(M + 1\) elements including the targeted element \(i\), is created alongside adjacent elements of \(i\). SBC is utilized because of lower computational cost and improved robustness, for more details please refer to \cite{30}. We employ \(M = 2K\) for improved robustness as mentioned in various previous studies \cite{30,32}, where \(K\) is number of coefficients in a polynomial:
\begin {equation}
K(r,d)=\frac{1}{d!} \prod_{l=1}^{d}(r+l),  
\end {equation}
where \(d\in [2,3]\) is space dimensions. This entire stencil is converted to reference space \(S^{'}\). The \(r_{th}\) order reconstructed polynomials are extended by a set of polynomial basis functions \(\phi _{k}(\xi ,\eta ,\zeta )\):
 \begin {equation}
p(\xi ,\eta ,\zeta)=\sum_{k=0}^{K} a_{k}  \phi _{k}(\xi ,\eta ,\zeta ) =U_{0}+\sum_{k=1}^{K} a_{k}  \phi _{k}(\xi ,\eta ,\zeta),
\end {equation}
 where \(U_{0}\) demonstrates cell average solution in specified cell \(i\), \(a_{k}\) denotes the degrees of freedom. To guarantee the limitation of \autoref{e4} for the element \(i\) and satisfy choices of all the elements in the stencil, the basis functions are expressed as
 \begin {equation}
\phi _{k}(\xi ,\eta ,\zeta ) =\psi _{k}(\xi ,\eta ,\zeta  )-\frac{1} {\mid V_{0}^{'}\mid\ } \int_{V_{0}^{'}}^{} \psi _{k}d\xi d\eta d\zeta \quad
   \psi _{k}=\xi ,\eta ,\zeta ,\xi^{2} ,\eta^{2} ,\zeta^{2} ,\xi \eta,\xi \zeta ,\eta \zeta ... \quad k=1,2...K.
  \end {equation}

The degrees of freedom \(a_{k}\) for cell \(m\) in the stencil are defined by meeting requirement that cell average value of reconstructed polynomial \(p(\xi ,\eta ,\zeta)\) maintain the same with the cell average value of solution \(U_{m}\).
 
  \begin {equation}
  \int_{V_{m}^{'} }^{} p(\xi ,\eta ,\zeta )d\xi d\eta d\zeta =\mid V_{m}^{'}\mid U_{0} +\sum_{k=1}^{K}\int_{V_{m}^{'} }^{} a_{k} \phi _{k}d\xi d\eta d\zeta =\mid V_{m}^{'}\mid U_{m}\quad  m=1,2...M.
 \end {equation}

Indicating the calculations of basis function integral \(k\) of cell \(m\) in stencils, the vector \(A_{mk}\) and \(b\) respectively is expressed as:
  \begin {equation}
A_{mk}=\int_{V_{m}^{'} }^{}  \phi _{k}d\xi d\eta d\zeta   \quad b_{m}=\mid V_{m}^{'}\mid (U_{m}-U_{0}).
     \end {equation}

The degrees of freedom in the equations can be formulated in matrix format
  \begin {equation}
  \sum_{k=1}^{K} A_{mk}a_{k} =b_{m}.
  \end {equation}

The matrix \(A_{mk}\) exclusively includes the geometric data for each element within the considered stencil \(S\). \(A_{mk}\) can be simplified by employing QR decomposition method with Householder transformations \cite{33}.

 \begin {equation}
 a_{k} =(A_{mk}^{T}A_{mk})^{-1}A_{mk}^{T}b_{m}=((QR)^{T}(QR))^{-1}A_{mk}^{T}b_{m}=(R^{T}R)^{-1}A_{mk}^{T}b_{m}.
 \end {equation}

\subsection{WENO schemes}
\label{}

In hyperbolic systems, high-order linear reconstruction scheme is inappropriate to handle discontinuous shockwaves. WENO scheme utilizes a combination of non-linear reconstruction polynomials obtained by central stencil and directional stencil \cite{5}. The weight assigned to each polynomial depends on solution's smoothness. The polynomials employed in WENO are stated as follows:
 \begin {equation}
p_{i}(\xi ,\eta ,\zeta )^{WENO}=\sum_{s=1}^{s_{t}} \omega _{s}p_{s}(\xi ,\eta ,\zeta ),
 \end {equation}
 where \(s_{t}\) is total number of stencils, nonlinear weight \(w_{k}\) is given by
 \begin {equation}
\omega _{s}=\frac{\tilde{\omega }  }{ \sum_{s=1}^{s_{t}} \tilde{\omega }} \quad ,where \quad \tilde{\omega }=\frac{\lambda _{s}}{(\epsilon +SI_{s})^{b}}.
 \end {equation}

Smoothness indicator \(SI_{s}\) is calculated by:
\begin {equation}
SI_{s}=\sum_{q=1}^{r} \int_{V_{0}^{'}}^{} (D^{q}p_{s}(\xi ,\eta ,\zeta ))^{2}d\xi d\eta d\zeta ,
 \end {equation}
where \(r\) is the order of polynomial, \(b = 4\) and \(D \) is derivative operator, \(\epsilon\) is utilized to prevent division by zero, \(\lambda_{s}\) is linear weight. The central stencil is given significant large \(\lambda_{1}\) and directional stencils are given \(\lambda_{s} = 1\).  Previous subsection discussed the assembly of linear candidate stencil reconstructions, defined by \(p_{s}(\xi ,\eta ,\zeta )\), that exhibit precise orders. These reconstructed candidates maintain criteria of matching cell average results of the relevant elements. They are obtained by employing the identical limited least-squares techniques to resolve over-determined linear systems. While the application of characteristic varibles decomposition in process substantially raises the computational footprints of schemes, numerous studies have consistently demonstrated the superior performance of WENO reconstructions in suppressing numerical oscillations when applied to characteristic variables, compared to primitive or conservative variables. This is particularly pronounced in cases featuring strong shocks and contact discontinuities, as evidenced by \cite{5,15}.

\subsection{CWENO schemes}
\label{}

\begin{figure*}
	\centering

     \subfigure[]{
    \includegraphics[scale=0.3]{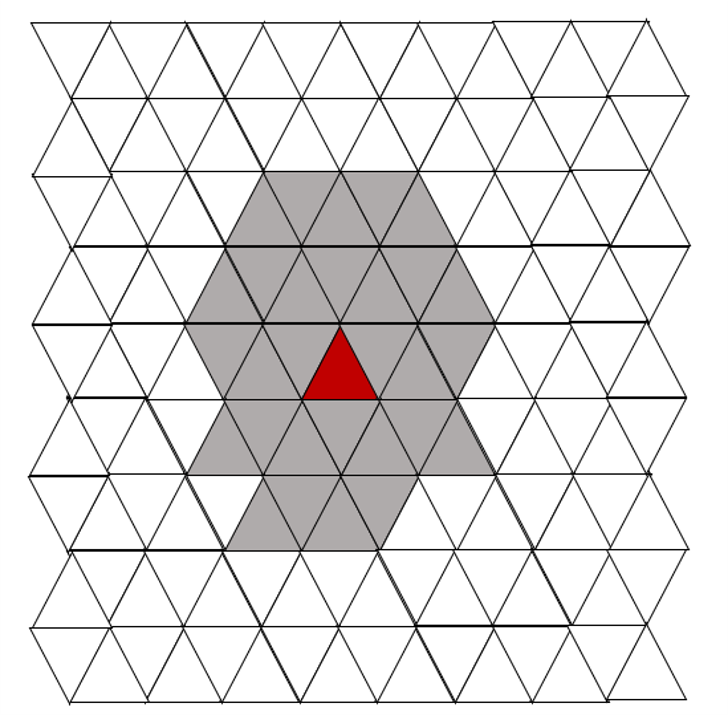}}
    \subfigure[]{
    \includegraphics[scale=0.3]{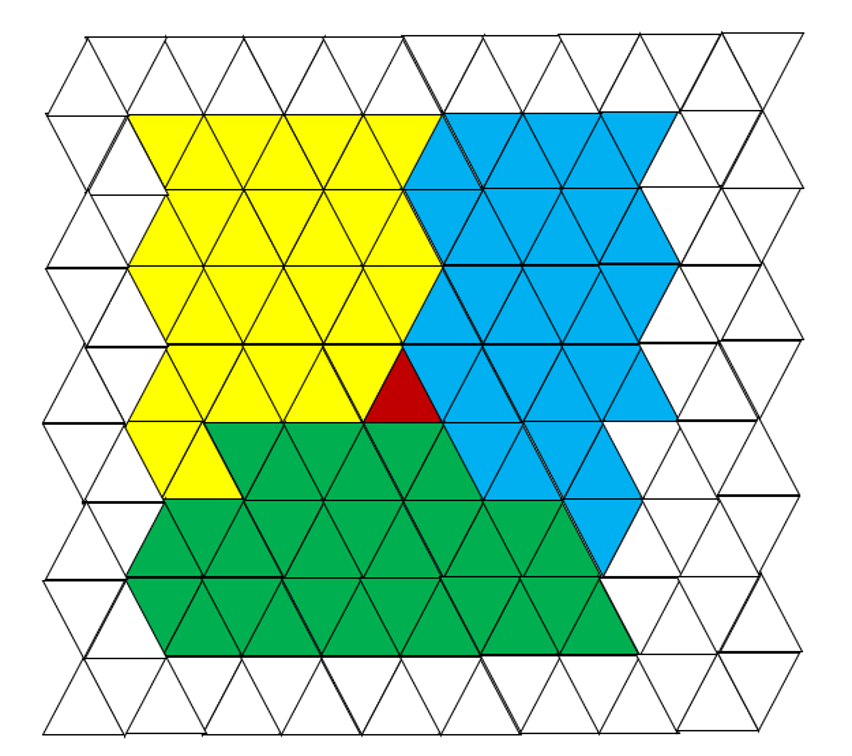}}
    \subfigure[]{
    \includegraphics[scale=0.4]{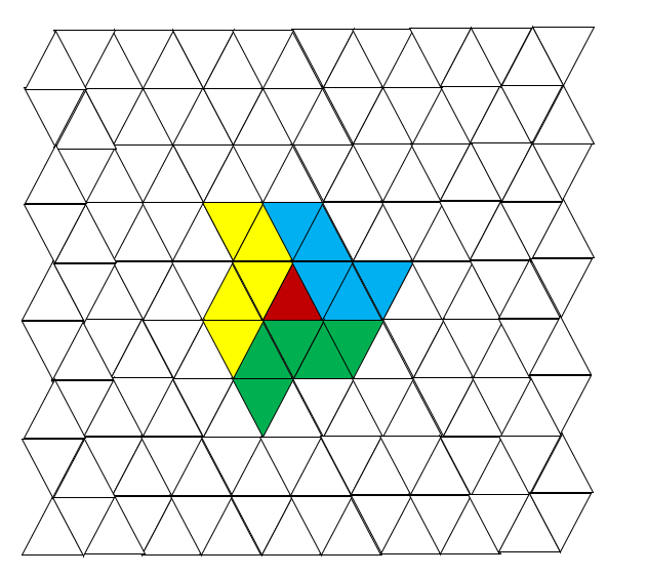}}
      
	\caption{The stencil for WENO on triangular grid is the fifth-order central stencil(a) and fifth-order directional stencils(b). The stencil for CWENO on triangular grid is the fifth-order central stencil(a) and third-order small directional stencils(c). The considered cell displayed in red, while the central stencil elements are displayed in gray, and each directional stencil is indicated by different colors.}
	\label{fig1}
\end{figure*}

The CWENO scheme originates from the methodology proposed by Tsoutsanis and Dumbser \cite{8}. This approach utilizes a combination of optimal polynomial \(p_{opt}\) with high-order and a set of polynomials with lower-order. The high-order polynomial is reconstructed by employing an extended central stencil that corresponds to the required polynomial order, while lower-order polynomials are reconstructed using directional stencils with compactness. All polynomials must fullfil same condition of matching the cell average values of the results, and solve with the same limited least-squares technique. In comparison to traditional WENO schemes, the CWENO scheme decreases computational expenses by employing smaller directional stencils within the central stencil (see \autoref{fig1}). This method establishes an optimal polynomial, which is a high-order polynomial that satisfies the desired accuracy order. The optimal polynomial's definition is provided by

\begin {equation}
p_{opt}=\sum_{s=1}^{s_{t}} \lambda _{s}p_{s}(\xi ,\eta ,\zeta),
\label{e16}
 \end {equation}
 where \(s_{t}\) is total number of stencils,  \(s=1\) is central stencil, \(s=(2,3,...s_{t})\) is directional stencil, and \(\lambda _{s}\) is linear coefficient, whose sum equals to \(1\). The \(p_{1}\) obtained by deducting lower-order polynomials from optimal polynomial is given in the following manner:
\begin {equation}
 p_{1}(\xi ,\eta ,\zeta )=\frac{1}{\lambda _{1}}( p_{opt}(\xi ,\eta ,\zeta )-\sum_{s=2}^{s_{t}} \lambda _{s}p_{s}(\xi ,\eta ,\zeta )).
 \label{e17}
 \end {equation}

The CWENO reconstructed polynomial is expressed as nonlinear combination of all polynomials by:

\begin {equation}
p_{i}(\xi ,\eta ,\zeta )^{CWENO}=\sum_{s=1}^{s_{t}} \omega _{s}p_{s}(\xi ,\eta ,\zeta ),
 \end {equation}
 where \(\omega _{s}\) is nonlinear weight and can be calculated the same with WENO. Unlike the WENO scheme, the linear coefficients \(\lambda _{s}\) are defined by
  \begin {equation} 
 \lambda _{1}=1-\frac{1}{\lambda_{1} ^{'} }  \quad and \quad  \lambda _{s}=\frac{1-\lambda _{1}}{s_{t} -1} ,
 \label{e18}
 \end {equation}
where \(\lambda_{1} ^{'}\) is an arbitrary value. When the scale of local flow is smooth, high-order reconstruction gradually approaches restoration to \(p_{1}\). As discontinuities are approached, the influence of large stencil diminishes, and smooth small directional stencils take control over the polynomial of final reconstruction. Consequently, CWENO's precision relies on reconstruction applied to large stencil, while reconstruction effectiveness of non-smooth areas is determined by small stencils.

While compactness of CWENO schemes has its benefits in regions with discontinuities on a coarse grid, the nonlinear weight of central stencil inaccurately decrease in smooth areas of the flow as a result of grid topology or morphology. As a consequence, this characteristic leads to substantial reduction in accuracy order compared with traditional WENO schemes \cite{30}. The outcomes achieved with CWENO schemes demonstrate a greater reliance on the parameters involved, due to the need for a delicate balance and adjustment between the higher-order approximation obtained from central stencil and directional stencils.

\subsection{TENO schemes}

Similar to previous CWENO scheme, the recently proposed TENO scheme in unstructured grids aims to create stencils comprising a significant central stencil and various directional small stencils \cite{28}. TENO schemes adopt high-order reconstruction techniques that utilizes large central stencil to resolve smooth region, while employing a serious of schemes with the small directional stencils to capture discontinuities. In order to differentiate between discontinuities and smoother scales, the TENO weighting approach utilizes a core algorithm that evaluates candidate stencils's smoothness with noticeable separation of scale \cite{15,34}. This is achieved by assessing the scale separation of a stencil using the following definition.

\begin {equation}
 \gamma _{s}=\frac{1}{(SI_{s}+\epsilon )^{6}} \quad s=1,2...s_{t}.
\end {equation}

The definition of \(SI_{s}\) is defined the same with WENO scheme. In comparison to WENO and CWENO schemes, for better scale separation and lower dissipation property, a larger exponent parameter is considered \cite{5,27}. It should be noted that previous studies using Cartesian TENO schemes \cite{34} have shown the effectiveness of the proposed approach in achieving good scale separation. In TENO, the ENO-like method is used in stencil selection to assess each stencil as either a smooth or non-smooth candidate. To start, the indicator of measured smoothness is defined as:
\begin {equation}
\chi _{s} =\frac{\gamma _{s}}{\sum_{s=1}^{s_{t}} \gamma _{s}}, 
\label{e22}
\end {equation}
and then exposed to a rigorous truncation function,
\begin{equation}
\delta _{s}=\begin{cases} 0, \text{ if } \chi _{s}< C_{T} \\ 1, \text{ otherwise }\end{cases}
\label{e23}
\end{equation}
where  \(C_{T}\) determines wavenumber and acts as a boundary between the smooth and nonsmooth scales. In practical applications, the selection of \(C_{T}\) is quite flexible, ranging from \(10^{-5}\) to \(10^{-7}\). The proposed approach has been proven to possess numerical robustness and low-dissipation characteristics through extensive simulations performed on broad flow scales \cite{15,35}. In stencil selection steps, the value of parameter \(C_{T}\) is set to \(10^{-6}\) in order to achieve optimal performance. Further optimization related to the choice of \(C_{T}\) is not covered in this paper. For more detailed information, interested readers are recommended to consult the discussions in \cite{15,36}.

The large candidate stencil possesses the highest level of accuracy and excellent spectral properties, making it highly effective in resolving smooth flow. A collection of smaller directional stencils has been developed to capture potential discontinuities. These smaller stencils employ a non-linear adaptation technique to enhance their performance. As a result, the process of determining final reconstruction at cell interface is carried out in the following two distinct steps:

(i) Gather candidate stencils and utilize the ENO-type stencil selection method using \autoref{e22} and \autoref{e23}. In the event that largest stencil is continuous, reconstruction scheme is expressed as:
\begin {equation}
p_{i}(\xi ,\eta ,\zeta )^{TENO}=p_{1}(\xi ,\eta ,\zeta ).
 \end {equation}

(ii) If it is found that the largest candidate stencil is not smooth, the nonlinear adjustment between the smaller stencils will capture the discontinuities by ensuring the property and implementing the selection of ENO-type stencils. Consequently, the ultimate reconstruction scheme will be formulated by
\begin {equation}
p_{i}(\xi ,\eta ,\zeta )^{TENO}=\sum_{s=2}^{s_{t}} \omega _{s}p_{s}(\xi ,\eta ,\zeta ),
\quad where \quad
\omega _{s}=\frac{\delta _{s}}{\sum_{s=2}^{s_{t}}\delta _{s} } ,\quad s=2,3...s_{t}.
 \end {equation}

\subsection{CTENO schemes}
\label{}

Following the CWENO schemes, we utilize the optimal polynomial to calculate \(p_{1}\) in \autoref{e16} and \autoref{e17}. The linear coefficients \(\lambda _{s}\) are defined by \autoref{e18}. The reconstruction polynomials of CTENO schemes are as follows:
\begin {equation}
 p_{i}(\xi ,\eta ,\zeta )^{CTENO}=\frac{\omega _{1}}{\lambda _{1}} ( p_{opt}(\xi ,\eta ,\zeta )-\sum_{s=2}^{s_{t}} \lambda _{s}p_{s}(\xi ,\eta ,\zeta ))+\sum_{s=2}^{s_{t}} \omega _{s}p_{s}(\xi ,\eta ,\zeta )
\end {equation}

The stencil selection strategy is identical to CWENO and TENO, which contains a central stencil and a collection of directional stencils with small size. Unlike WENO and CWENO, TENO system does not calculate the non-linear coefficient \(\omega _{s}\) directly. In CTENO schemes, we apply the TENO weight strategy to evaluate and capture discontinuities. The scale separation strategy of stencils is the same with that used in TENO schemes \cite{28,29}. After judging the smooth and non-smooth stencil with ENO-like stencil selection strategy, ultimate reconstruction are described in two steps:

(i) If large stencil is continuous, reconstruction scheme is expressed as:
\begin {equation}
p_{i}(\xi ,\eta ,\zeta )^{CTENO}=p_{1}(\xi ,\eta ,\zeta ).
 \end {equation}

(ii) If large candidate stencil is non-smooth , nonlinear combination between the small candidate stencils will capture the discontinuities. Reconstruction scheme will be described by
\begin {equation}
p_{i}(\xi ,\eta ,\zeta )^{CTENO}=\sum_{s=2}^{s_{t}} \omega _{s}p_{s}(\xi ,\eta ,\zeta ),
\quad where \quad
\omega _{s}=\frac{\delta _{s}}{\sum_{s=2}^{s_{t}}\delta _{s} } ,\quad s=2,3...s_{t}.
 \end {equation}

One crucial feature of CTENO is integration of optimal polynomial with high-order based on large central stencil. In smooth area, optimal polynomial is obtained to achieve the acquired accuracy level, while lower-order polynomials include smooth data at the points of discontinuous, thus reducing oscillations in calculated results. All polynomials involved in stencils are bound by the requirements as specified, matching  the average value of solution in each cell. Taking into account previous version of unstructured TENO schemes \cite{28,29}, the CTENO scheme with a high-order optimal polynomial and low-order polynomials allows for construction of unstructured CTENO reconstructions with arbitrarily high orders and exceptional numerical stability, as demonstrated in the subsequent section. For the algorithm of CTENO scheme, outlined in \autoref{a1}.

\begin{algorithm}[H]
\SetAlgoLined
\caption{Pseudocode for CTENO schemes computation for cell i}
\label{a1}
PROCEDURE CTENO scheme\;
Compute the optimal polynomial \(p_{1}\)\;
\For{each admissible stencil $s$}
{
    Compute the Smooth Indicator $SI_{s}$\; 
}
!Compute Weight\;
\For{each variable j}{
   $\omega _{s}= 0$\; 
} 
\For{each admissible stencil $s$}
{
    Compute $\gamma _{s}=\frac{1}{(SI_{s}+\epsilon )^{6}}$\; 
}

Computer sum $\sum_{s=1}^{s_{t}}\gamma _{s}$\;
\For{each admissible stencil $s$}
{
    Compute $\chi _{s} =\frac{\gamma _{s}}{\sum_{s=1}^{s_{t}} \gamma _{s}}$\;
}
\For{each admissible stencil $s$}
{
    Compute $\delta _{s}=\begin{cases} 0, & \text{ if } \chi _{s}< C_{T} \\ 1, & \text{ otherwise }\end{cases}$\;
}

\eIf{$\delta _{1}=1$}
{
    $\omega _{1}=1$\;
    \For{other admissible stencil $s$}
    {
     $\omega _{s}=0$
    }
    
}
{
    $\omega _{1}=0$\;
    \For{other admissible stencil $s$}
    {
     $\omega _{s}=\frac{\delta _{s}}{\sum_{s=2}^{s_{t}} \delta _{s}}$
    } 
}
\end{algorithm}

\subsection{CTENOZ schemes}
The CTENOZ essentially follows the framework of CTENO, the determinations of stencil selection strategy and optimal polynomial remains unchanged as CTENO. The primary distinction lies in the implementation of the scale separation strategy. The smoothness indicators of CWENO and CTENO schemes are obtained with polynomials from different orders. The WENOZ scheme, initially proposed by Borges et al. and Castro et al. \cite{7,55}, is considered in the research. However, it is customized to handle polynomials with unequal orders, reconstructed stencils of varying sizes, and arbitrary shape of cells, as reported in recent studies by \cite{56,57}. The CTENOZ weighting strategy for scale separation aims to assess the degree of smoothness in candidate stencils by:
\begin {equation}
 \gamma _{s}=1+\frac{\tau }{(SI_{s}+\epsilon )} \quad s=1,2...s_{t}.
\end {equation}

 \(\tau\) is universal indicator of oscillation, which is regarded as the absolute difference values of smoothness indicators, described by:

\begin {equation}
\tau =(\frac{\sum_{s=2}^{s_{t}}\mid SI_{s}-SI_{1}\mid  }{s_{t}-1} )^{6}
\end {equation}

After measuring the smoothness with strong scale-separation formula of CTENOZ schemes, the reconstruction polynomial will be calculated the same with TENO and CTENO schemes.

(i) If it is determined that central stencil is smooth, final reconstruction on the cell interface will be provided as
\begin {equation}
p_{i}(\xi ,\eta ,\zeta )^{CTENOZ}=p_{1}(\xi ,\eta ,\zeta ).
 \end {equation}

(ii) Otherwise, the ultimate reconstruction scheme will be formulated by
\begin {equation}
p_{i}(\xi ,\eta ,\zeta )^{CTENOZ}=\sum_{s=2}^{s_{t}} \omega _{s}p_{s}(\xi ,\eta ,\zeta ),
\quad where \quad
\omega _{s}=\frac{\delta _{s}}{\sum_{s=2}^{s_{t}}\delta _{s} } ,\quad s=2,3...s_{t}.
 \end {equation}

\begin{figure*}[h]
	\centering
    \includegraphics[scale=0.4]{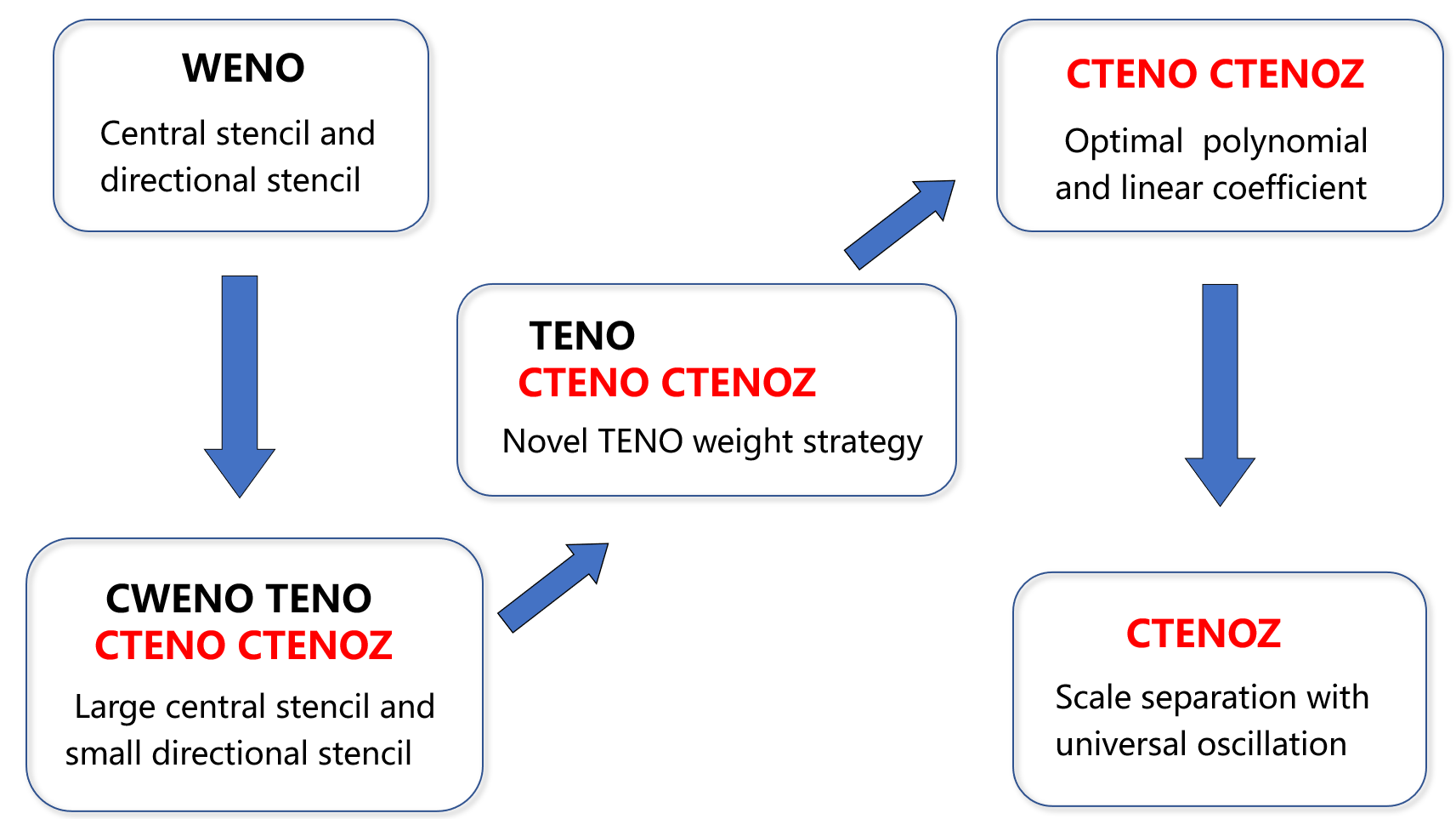}
	\caption{The categorized features for WENO, CWENO, TENO, CTENO, and CTENOZ schemes are listed. The CTENO and CTENOZ shemes developed in this paper are marked in red.}
	\label{mindmap}
\end{figure*}

The algorithm of CTENOZ schemes is shown in the \autoref{a2}. Comparison between WENO, CWENO, TENO, CTENO, and CTENOZ schemes are shown in \autoref{mindmap}. The stencil selection, weight strategy, polynomial, linear coefficient, and scale separation for the schemes are categorized.

\begin{algorithm}[H]
\SetAlgoLined
\caption{Pseudocode for CTENOZ schemes computation for cell i}
\label{a2}
PROCEDURE CTENOZ scheme\;
Compute the optimal polynomial \(p_{1}\)\;
\For{each admissible stencil $s$}
{
    Compute the Smooth Indicator $SI_{s}$\; 
}

Compute universal indicator of oscillation\;
\(\tau =(\frac{\sum_{s=2}^{s_{t}}\mid SI_{s}-SI_{1}\mid  }{s_{t}-1} )^{6}\) \;

!Compute Weight\;
\For{each variable j}{
   $\omega _{s}= 0$\; 
} 
\For{each admissible stencil $s$}
{
    Compute $ \gamma _{s}=1+\frac{\tau }{(SI_{s}+\epsilon )}$\; 
}

Computer sum $\sum_{s=1}^{s_{t}}\gamma _{s}$\;
\For{each admissible stencil $s$}
{
    Compute $\chi _{s} =\frac{\gamma _{s}}{\sum_{s=1}^{s_{t}} \gamma _{s}}$\;
}
\For{each admissible stencil $s$}
{
    Compute $\delta _{s}=\begin{cases} 0, & \text{ if } \chi _{s}< C_{T} \\ 1, & \text{ otherwise }\end{cases}$\;
}

\eIf{$\delta _{1}=1$}
{
    $\omega _{1}=1$\;
    \For{other admissible stencil $s$}
    {
     $\omega _{s}=0$
    }
    
}
{
    $\omega _{1}=0$\;
    \For{other admissible stencil $s$}
    {
     $\omega _{s}=\frac{\delta _{s}}{\sum_{s=2}^{s_{t}} \delta _{s}}$
    } 
}
\end{algorithm}

\subsection{Fluxes approximation}
\label{}

Harten-Lax-van Leer-Contact Riemann solver(HLLC) \cite{37} is applied to inviscid fluxes, providing reliable approximations. The discontinuous left and right state $\mathbf{U}_{L}$ , $\mathbf{U}_{R}$ is

\begin {equation}
\mathbf{U} =\begin{cases}
\mathbf{U}_{L}, \text{ if } S_{L}> 0 \\
\mathbf{U}_{L}^{*}, \text{ if } S_{L}\le  0< S_{M}\\
\mathbf{U}_{R}^{*}, \text{ if } S_{M}\le  0< S_{R} \\
  \mathbf{U}_{R}, \  \text{ if } S_{R}< 0
\end{cases}
 \end {equation}

and the corresponding flux is

\begin {equation}
 \mathbf{F}(\mathbf{U}_{L},\mathbf{U}_{R}) =\begin{cases}
\mathbf{F}_{L}, \text{ if } S_{L}> 0 \\
\mathbf{F}_{L}^{*}=\mathbf{F}_{L}+S_{L}(\mathbf{U}_{L}^{*}-\mathbf{U}_{L}), \text{ if } S_{L}\le  0< S_{M}\\
\mathbf{F}_{R}^{*}=\mathbf{F}_{R}+S_{R}(\mathbf{U}_{R}^{*}-\mathbf{U}_{R}), \text{ if } S_{M}\le  0< S_{R} \\
  \mathbf{F}_{R}. \  \text{ if } S_{R}< 0
\end{cases}
 \end {equation}

where $\mathbf{F}_{L}= \mathbf{F}(\mathbf{U}_{L})$ and $\mathbf{F}_{L}= \mathbf{F}(\mathbf{U}_{L})$. The acoustic wave-speed are evaluated by

\begin {equation}
 S_{L}=min(u_{L}-c_{L},\tilde{u} -\tilde{c} ),\quad S_{R}=max(u_{R}+c_{R},\tilde{u} +\tilde{c} ),
  \end {equation}

where
  
\begin {equation}
 \begin{cases}
D_{\rho }=\sqrt{\frac{\rho _{R}}{\rho _{L}} },\\
\tilde{u}=\frac{u_{L}+u_{R}D_{\rho }}{1+D_{\rho }}  ,  \\
\tilde{H}=\frac{H_{L}+H_{R}D_{\rho }}{1+D_{\rho }},  \\
 \tilde{c}= \sqrt{(\gamma -1)[\tilde{H}-\frac{1}{2}\tilde{u}^{2} ]}. \
\end{cases}
  \end {equation}

In order to determine the state $\mathbf{U}_{L}^{*}$ and $\mathbf{U}_{R}^{*}$, it is assumed that

\begin {equation}
  S_{M}=u^{*}=u_{L}^{*}=u_{R}^{*}, \quad p^{*}=p_{L}^{*}=p_{R}^{*}.
  \end {equation}

The speed of the contact wave is calculated as

\begin {equation}
S_{M}=\frac{\rho _{R}u_{R}(S_{R}-u_{R})-\rho _{L}u_{L}(S_{L}-u_{L})+p_{L}-p_{R}}{\rho _{R}(S_{R}-u_{R})-\rho _{L}(S_{L}-u_{L})},\quad p^{*}=\rho _{L}(u_{L}-S_{L})(u_{L}-S_{M})+p_{L},
  \end {equation}

where

\begin {equation}
 \begin{cases}
\rho_{L}^{*}=\rho_{L} \frac{S_{L}-u_{L}}{S_{L}-S_{M}} ,\rho_{R}^{*}=\rho_{R} \frac{S_{R}-u_{R}}{S_{R}-S_{M}},\\
\rho_{L}^{*}u_{L}^{*}=\frac{(S_{L}-u_{L})\rho _{L}u_{L}+p^{*}-p_{L}}{S_{L}-S_{M}}  ,\rho_{R}^{*}u_{R}^{*}=\frac{(S_{R}-u_{R})\rho _{R}u_{R}+p^{*}-p_{R}}{S_{R}-S_{M}},   \\
\rho_{L}^{*}v_{L}^{*}=\frac{(S_{L}-u_{L})\rho _{L}v_{L}}{S_{L}-S_{M}}, \rho_{R}^{*}v_{R}^{*}=\frac{(S_{R}-u_{R})\rho _{R}v_{R}}{S_{R}-S_{M}}, \\
\rho_{L}^{*}w_{L}^{*}=\frac{(S_{L}-u_{L})\rho _{L}w_{L}}{S_{L}-S_{M}}, \rho_{R}^{*}w_{R}^{*}=\frac{(S_{R}-u_{R})\rho _{R}w_{R}}{S_{R}-S_{M}}, \\
 E_{L}^{*}=\frac{(S_{L}-u_{L})E_{L}-p_{L}u_{L}+p^{*}S_{M}}{S_{L}-S_{M}},  E_{R}^{*}=\frac{(S_{R}-u_{R})E_{R}-p_{R}u_{R}+p^{*}S_{M}}{S_{R}-S_{M}}. \
\end{cases}
  \end {equation}

In the subsequent section, the HLLC method is utilized as the Riemann solver, which has been proven to possess robustness and reliability for the Euler equations.

\section{Numerical tests}
\label{}

The effectiveness of proposed CTENO and CTENOZ schemes will be assessed through a set of numerical simulations. We perform various benchmark test problems that involve both smooth and discontinuous solutions. The evaluation focuses on the performance of the CTENO and CTENOZ scheme in capturing interfaces in the inviscid compressible Euler equations. Several test problems are listed below: 
\begin{itemize}
  \item[\textbullet] Numerical Accuracy: The accuracy of the methods are tested using the two-dimensional Euler equations. In addition, the shock-density wave interaction is tested to evaluate the non-oscillatory characteristics. The impact of the linear weight of the central stencil $\lambda^{'}_{1}$ is detailed analyzed.
  \item[\textbullet] Shock-tube problem: This test examines the effectiveness of proposed schemes by comparing their solutions to the exact Sod problem, demonstrating their capability to accurately capture discontinuities.
  \item[\textbullet]2D Double Mach Reflection of a Strong Shock: This test problem evaluates the ability to capture discontinuities and the properties of dissipation. The refined mesh is considered and the computational cost is compared among different schemes. The parallel scalability is tested on the refined mesh. 
  \item[\textbullet]Single-Material Triple Point Problem: This test problem assesses the efficiency in capturing interfacial instability and smaller-scale structure. The computational cost associated with different orders is also investigated.
  \item[\textbullet]Kelvin-Helmholtz instability: This problem serves as a test to evaluate the precision and dissipation characteristics of numerical techniques. The dissipation of total kinetic energy is examined to assess the performance of schemes. 
  \item[\textbullet]Interaction of a Shock Wave with a Wedge in 2D: This test problem is used to assess the performance involving strong gradients that interact with vortices, and compare it  with the experimental results.
  \item[\textbullet]Helium Bubble Shock Wave: This test problem is utilized to evaluate the capability of simulating multicomponent flow with hybrid grid. The results is compared to experiment, revealing the interaction between shock waves and bubbles.
 \item[\textbullet]3D Explosion Problem: This problem is utilized for 3D tests. The simulation results provided by unstructured tetrahedral and hybrid elements were compared with exact solution.
  \end{itemize}

\subsection{Numerical Accuracy}
\label{}

\subsubsection{order tests for the two-dimensional Euler equations}

We consider the two-dimensional Euler equations, the initial conditions of the test are \(\rho (x,y,0)=1+0.2sin(2\pi (x+y))\), \(u(x,y,0)=1\), \(v(x,y,0)=1\), \(p(x,y,0)=1\). The computational domain is  \([0, 1] \times [0, 1]\), and periodic boundary conditions are employed. The exact solution of the density is  \(\rho (x,y,t)=1+0.2sin(2\pi (x+y-2t))\). The numerical errors  \(L^{2}error\) and \(L^{\infty}error\) are calculated by:

\begin {equation}
L^{2}error=\sqrt{\frac{\sum_{i}^{}\int_{V_{i}}^{}(U_{e}(x,t)-U_{c}(x,t))^{2}dV  }{\sum_{i}^{} \mid V_{i} \mid  } },
\end {equation}

\begin {equation}
L^{\infty}error =Max\mid U_{e}(x,t)-U_{c}(x,t)\mid ,
\end {equation}

where \(U_{e}(x,t)\) and \(U_{c}(x,t)\) are the exact and numerical results. The errors and orders of accuracy at time  \(t=1\) are displayed in \autoref{ordercteno4}, \autoref{orderctenoz4} and \autoref{order15}. We select central stencil linear weight of $\lambda_{1} ^{'}=10^{4}$ and $\lambda_{1} ^{'}=10^{15}$ to examine the capabilities of high order polynomials, as demonstrated in \cite{8,28}. It is observed that the convergence rates are significantly influenced by the value of $\lambda_{1} ^{'}$. When $\lambda_{1} ^{'}=10^{4}$, as shown in \autoref{ordercteno4}, the spatial accuracy is deteriorated in CTENO scheme for all the orders of discretization and mesh resolutions, which is similar to the test results for CWENO scheme \cite{8}. From the previous tests \cite{8}, the spatial accuracy in CWENO scheme is deteriorated when $\lambda_{1} ^{'}$ was set to $10^{3}$ and $10^{7}$. However, employing an excessively large $\lambda_{1} ^{'}$ value of $10^{15}$ is necessary to attain the correct order. This necessity arises from the interplay of smoothness indicators associated with polynomials of varying accuracy levels, which can hinder CWENO schemes from realizing their full potential. An extraordinarily large $\lambda_{1} ^{'}$ is required for marching high order polynomials. This issue also applies in the present CTENO scheme.

From \autoref{orderctenoz4}, the CTENOZ scheme achieves convergence rates close to the theoretical orders within 5th order of accuracy when $\lambda_{1} ^{'}=10^{4}$, but fails in refined mesh to achieve 6th and 7th spacial order. Note that the universal oscillation indicators $\tau$ is proposed in CTENOZ which arise from polynomials of different orders. The introduction of $\tau$ makes CTENOZ scheme much less sensitive to $\lambda_{1} ^{'}$, however, the influence of low order directional stencil prevent the scheme to achieve high order at $\lambda_{1} ^{'}=10^{4}$. When $\lambda_{1} ^{'}=10^{15}$ is used, as shown in \autoref{order15}, both CTENO and CTENOZ schemes achieve the theoretical order of accuracy. Under this extremely large $\lambda_{1} ^{'}$, the CTENO and CTENOZ schemes provide the same errors and accuracy orders due to the choice of large center stencil in smooth areas. In the following subsection, the effect of $\lambda_{1} ^{'}$ on the performance of different schemes are further discussed.

\begin{table}[h]
\centering
\caption{Errors and accuracy with two dimensional Euler equations for CTENO scheme with $\lambda_{1} ^{'}=10^{4}$. \(h\) is the number of edges. \(Ne\) is the total number of elements. Both \(L^{\infty}error\) and \(L^{2}error\) are calculated based on \(h\) and \(Ne\).}
\begin{tabular}{ccccccc}
\toprule
 &\(h\) & \(Ne\) & \(L^{\infty}error\) &\(L^{\infty}order\) & \(L^{2}error\) & \(L^{2}order\) \\
\midrule
CTENO3 & 1/10& 100  &  1.13E-01  &      & 7.96E-02 &       \\
       & 1/20& 400  &  2.42E-02  & 2.22 & 1.52E-02 & 2.39    \\
       & 1/40& 1600 &  3.88E-03  & 2.64 & 1.99E-03 & 2.94   \\
       & 1/80& 6400 &  8.92E-04  & 2.12 & 3.09E-04 & 2.69    \\
      & 1/160& 25600&  2.18E-04  & 2.03 & 6.17E-05 & 2.32    \\
CTENO4 & 1/10& 100  &  6.71E-02  &      & 4.82E-02 &       \\
       & 1/20& 400  &  8.45E-03  & 2.99 & 5.48E-03 & 3.14    \\
       & 1/40& 1600 &  2.01E-03  & 2.07 & 7.65E-04 & 2.84   \\
       & 1/80& 6400 &  8.14E-04  & 1.31 & 2.70E-04 & 1.50    \\
      & 1/160& 25600&  2.93E-04  & 1.48 & 7.82E-05 & 1.79    \\
CTENO5 & 1/10& 100  &  5.36E-02  &      & 3.45E-02 &       \\
       & 1/20& 400  &  1.03E-02  & 2.38 & 4.49E-03 & 2.94    \\
       & 1/40& 1600 &  2.40E-03  & 2.10 & 8.76E-04 & 2.36   \\
       & 1/80& 6400 &  1.15E-03  & 1.06 & 3.49E-04 & 1.33    \\
      & 1/160& 25600&  4.51E-04  & 1.35 & 1.05E-04 & 1.74    \\
CTENO6 & 1/10& 100  &  5.68E-02  &      & 3.66E-02 &       \\
       & 1/20& 400  &  1.12E-02  & 2.34 & 5.13E-03 & 2.83    \\
       & 1/40& 1600 &  3.02E-03  & 1.89 & 1.02E-03 & 2.33   \\
       & 1/80& 6400 &  1.26E-03  & 1.26 & 3.56E-04 & 1.52    \\
      & 1/160& 25600&  4.01E-04  & 1.65 & 8.41E-05 & 2.08    \\
CTENO7 & 1/10& 100  &  7.77E-02  &      & 4.71E-02 &       \\
       & 1/20& 400  &  2.92E-02  & 1.41 & 1.20E-02 & 1.98    \\
       & 1/40& 1600 &  6.77E-03  & 2.11 & 2.06E-03 & 2.54   \\
       & 1/80& 6400 &  6.22E-03  & 0.12 & 2.09E-03 &-0.02    \\
      & 1/160& 25600&  2.30E-03  & 1.44 & 6.83E-04 & 1.61    \\
\bottomrule
                 
\end{tabular}
\label{ordercteno4}
\end{table}

\begin{table}[h]
\centering
\caption{ Errors and accuracy with two dimensional Euler equations for CTENOZ scheme with $\lambda_{1} ^{'}=10^{4}$. \(h\) is the number of edges. \(Ne\) is the total number of elements. Both \(L^{\infty}error\) and \(L^{2}error\) are calculated based on \(h\) and \(Ne\).}
\begin{tabular}{ccccccc}
\toprule
&\(h\) & \(Ne\) & \(L^{\infty}error\) &\(L^{\infty}order\) & \(L^{2}error\) & \(L^{2}order\) \\
\midrule
CTENOZ3 & 1/10& 100  &  1.06E-01  &      & 7.54E-02 &       \\
        & 1/20& 400  &  1.97E-02  & 2.42 & 1.40E-02 & 2.43    \\
        & 1/40& 1600 &  2.65E-03  & 2.90 & 1.88E-03 & 2.90   \\
        & 1/80& 6400 &  3.35E-04  & 2.98 & 2.37E-04 & 2.98    \\
       & 1/160& 25600&  4.20E-05  & 3.00 & 2.97E-05 & 3.00    \\
CTENOZ4 & 1/10& 100  &  5.39E-02  &      & 3.72E-02 &       \\
        & 1/20& 400  &  4.32E-03  & 3.64 & 2.96E-03 & 3.65    \\
        & 1/40& 1600 &  2.80E-04  & 3.94 & 1.95E-04 & 3.92   \\
        & 1/80& 6400 &  1.77E-05  & 3.99 & 1.24E-05 & 3.98    \\
       & 1/160& 25600&  1.10E-06  & 4.00 & 7.77E-07 & 3.99    \\
CTENOZ5 & 1/10& 100  &  2.31E-02  &      & 1.63E-02 &       \\
        & 1/20& 400  &  8.68E-04  & 4.73 & 6.18E-04 & 4.72    \\
        & 1/40& 1600 &  2.83E-05  & 4.94 & 2.00E-05 & 4.95   \\
        & 1/80& 6400 &  8.87E-07  & 4.99 & 6.25E-07 & 5.00    \\
       & 1/160& 25600&  2.77E-08  & 5.00 & 1.92E-08 & 5.02    \\
CTENOZ6 & 1/10& 100  &  1.33E-02  &      & 9.11E-03 &       \\
        & 1/20& 400  &  2.49E-04  & 5.74 & 1.71E-04 & 5.74    \\
        & 1/40& 1600 &  3.82E-06  & 6.03 & 2.67E-06 & 6.00   \\
        & 1/80& 6400 &  4.35E-08  & 6.46 & 2.98E-08 & 6.48    \\
       & 1/160& 25600&  3.33E-09  & 3.71 & 2.18E-09 & 3.77    \\
CTENOZ7 & 1/10& 100  &  9.51E-03  &      & 6.62E-03 &       \\
        & 1/20& 400  &  6.82E-05  & 7.12 & 4.83E-05 & 7.10    \\
        & 1/40& 1600 &  5.04E-07  & 7.08 & 3.52E-07 & 7.10   \\
        & 1/80& 6400 &  1.94E-08  & 4.70 & 1.31E-08 & 4.74    \\
       & 1/160& 25600&  4.20E-09  & 2.21 & 2.82E-09 & 2.22    \\
		\bottomrule
		
	\end{tabular}
	\label{orderctenoz4}
\end{table}

\begin{table}[h]
\centering
\caption{ Errors and accuracy with two dimensional Euler equations for CTENO and CTENOZ schemes with $\lambda_{1} ^{'}=10^{15}$. \(h\) is the number of edges. \(Ne\) is the total number of elements. Both \(L^{\infty}error\) and \(L^{2}error\) are calculated based on \(h\) and \(Ne\).}
\begin{tabular}{ccccccc}
\toprule
&\(h\) & \(Ne\) & \(L^{\infty}error\) &\(L^{\infty}order\) & \(L^{2}error\) & \(L^{2}order\) \\
\midrule
CTENO3  & 1/10& 100  &  1.06E-01  &      & 7.54E-02 &       \\
CTENOZ3 & 1/20& 400  &  1.97E-02  & 2.42 & 1.40E-02 & 2.43    \\
        & 1/40& 1600 &  2.65E-03  & 2.90 & 1.88E-03 & 2.90   \\
        & 1/80& 6400 &  3.35E-04  & 2.98 & 2.37E-04 & 2.98    \\
       & 1/160& 25600&  4.20E-05  & 3.00 & 2.97E-05 & 3.00    \\
CTENO4  & 1/10& 100  &  5.39E-02  &      & 3.72E-02 &       \\
CTENOZ4 & 1/20& 400  &  4.32E-03  & 3.64 & 2.96E-03 & 3.65    \\
        & 1/40& 1600 &  2.80E-04  & 3.94 & 1.95E-04 & 3.92   \\
        & 1/80& 6400 &  1.77E-05  & 3.99 & 1.24E-05 & 3.97    \\
       & 1/160& 25600&  1.11E-06  & 4.00 & 7.80E-07 & 3.99    \\
CTENO5  & 1/10& 100  &  2.31E-02  &      & 1.63E-02 &       \\
CTENOZ5 & 1/20& 400  &  8.69E-04  & 4.73 & 6.18E-04 & 4.72    \\
        & 1/40& 1600 &  2.83E-05  & 4.94 & 2.00E-05 & 4.95   \\
        & 1/80& 6400 &  8.93E-07  & 4.99 & 6.32E-07 & 4.99    \\
       & 1/160& 25600&  2.80E-08  & 4.99 & 1.98E-08 & 4.99    \\
CTENO6  & 1/10& 100  &  1.33E-02  &      & 9.11E-03 &       \\
CTENOZ6 & 1/20& 400  &  2.49E-04  & 5.73 & 1.71E-04 & 5.73    \\
        & 1/40& 1600 &  3.90E-06  & 6.00 & 2.73E-06 & 5.97   \\
        & 1/80& 6400 &  6.03E-08  & 6.01 & 4.24E-08 & 6.01    \\
       & 1/160& 25600&  9.42E-10  & 6.00 & 6.61E-10 & 6.00    \\
CTENO7  & 1/10& 100  &  9.51E-03  &      & 6.63E-03 &       \\
CTENOZ7 & 1/20& 400  &  6.87E-05  & 7.11 & 4.87E-05 & 7.09    \\
        & 1/40& 1600 &  5.66E-07  & 6.92 & 4.00E-07 & 6.93   \\
        & 1/80& 6400 &  4.93E-09  & 6.84 & 3.48E-09 & 6.84    \\
       & 1/160& 25600&  4.39E-11  & 6.81 & 3.02E-11 & 6.85    \\

 \bottomrule
 \end{tabular}
  \label{order15}
\end{table}

\subsubsection{Shock-density wave interaction}
\label{}

The problem of shock-density wave interaction focus on a 1D shock wave with Mach number \(3\) and disturbed density field, as proposed in \cite{40}. This interaction results in the development of structures and discontinuities in small-scale. Consequently, this test problem is commonly utilized to assess abilities to capture shock waves and resolve wave numbers with different numerical methods. The influence of $\lambda_{1} ^{'}$ is also discussed in this problem. The initialization of flow field is

\begin {equation}
(\rho ,u,p)=\begin{cases}
  (3.8571,2.6294,10.333), \ -5\le x<-4 \\
  (1+0.2sin(5x),0,1). \  -4\le x\le 5
\end{cases}
 \end {equation}

The computational domain is \([-5, 5]\times [0, 2]\) and consists of 18282 triangular mesh cells distributed in a uniform manner, approximately with \(h \approx 1/20\). The time of the ultimate evolution is \(t = 1.8\). To obtain an accurate reference solution, a 1D Euler equations is utilized, with \(2000\) grid nodes and WENO scheme of 5th order.

\begin{figure*}[h]
	\centering
 
   \subfigure[]{
    \label{}
    \includegraphics[scale=0.4]{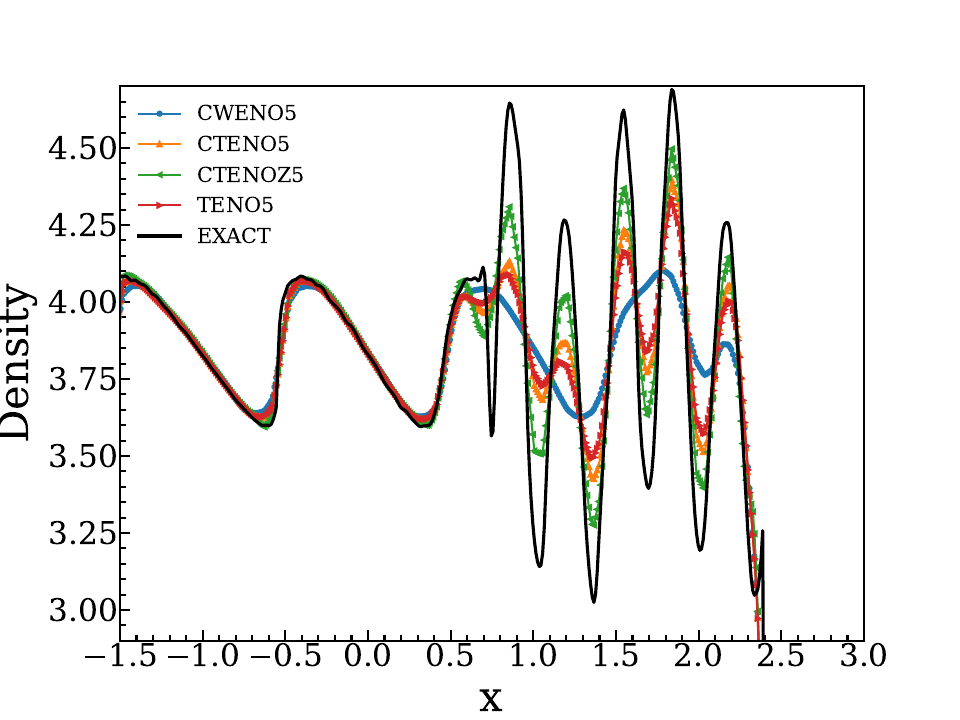}}
    \subfigure[]{
    \label{}
    \includegraphics[scale=0.4]{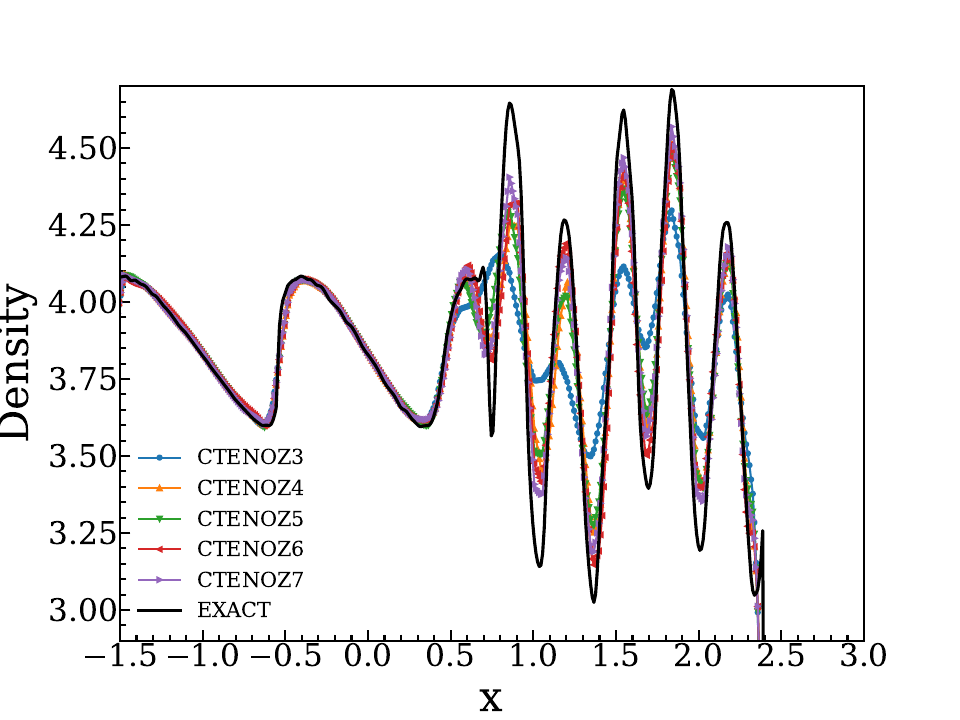}}
   
	\caption{shock-density wave interaction: solutions from (a) CWENO, TENO ,CTENO and CTENOZ of 5th-order,(b) CTENOZ scheme of various orders. Density profiles are compared with reference solution.}
	\label{fig4}
\end{figure*}

From the order test results, it is found that $\lambda_{1} ^{'}=10^{4}$ is applicable for CTENOZ scheme to achieve 5th spatial order for all mesh resolution and 7th order for coarse grid. Here, $\lambda_{1} ^{'}=10^{4}$ is also used to compare the resolution of different schemes. \autoref{fig4}(a) illustrates the successful capture of sound waves and small shocklets with robustness in various schemes. Regarding the entropy wave in the same accuracy order, the CTENO and CTENOZ schemes display a notable enhancement in comparison with CWENO and TENO. This improvement is attributed to their superior preservation of the density wave amplitude. The CWENO demonstrate highest dissipation, resulting in dissipation of small-scale structures. Among all schemes presented by \autoref{fig4}(a), CTENOZ schemes shows the least dissipation and better performance than other schemes under $\lambda_{1} ^{'}=10^{4}$. The improved performance with increasing order of accuracy in CTENOZ schemes is observed in \autoref{fig4}(b). 

\begin{figure*}[h]
	\centering
 
   \subfigure[]{
    \label{}
    \includegraphics[scale=0.4]{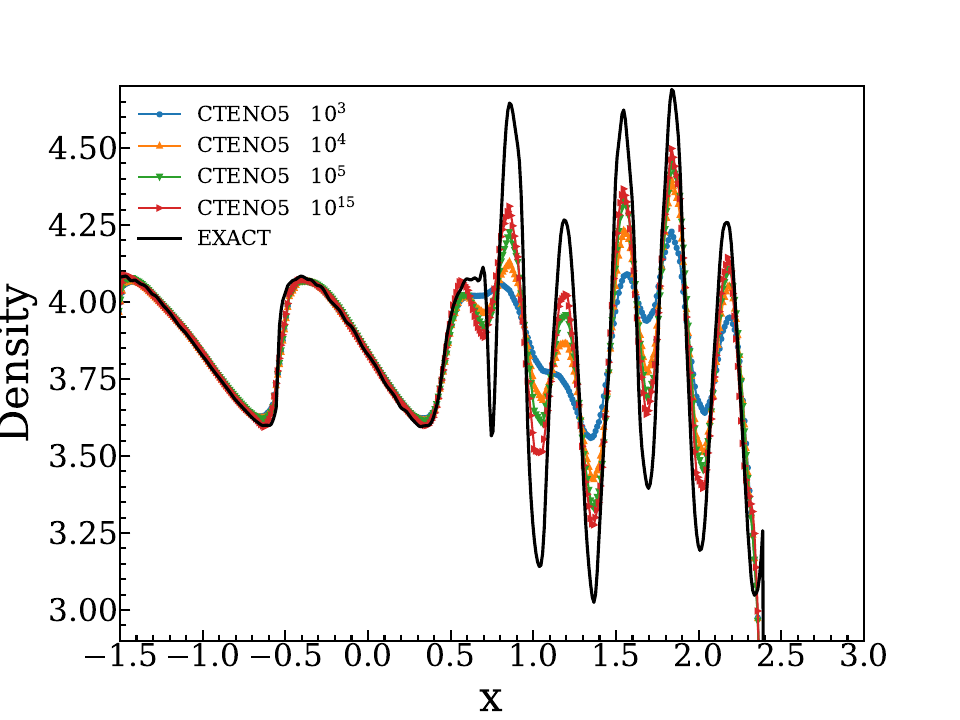}}
    \subfigure[]{
    \label{}
    \includegraphics[scale=0.4]{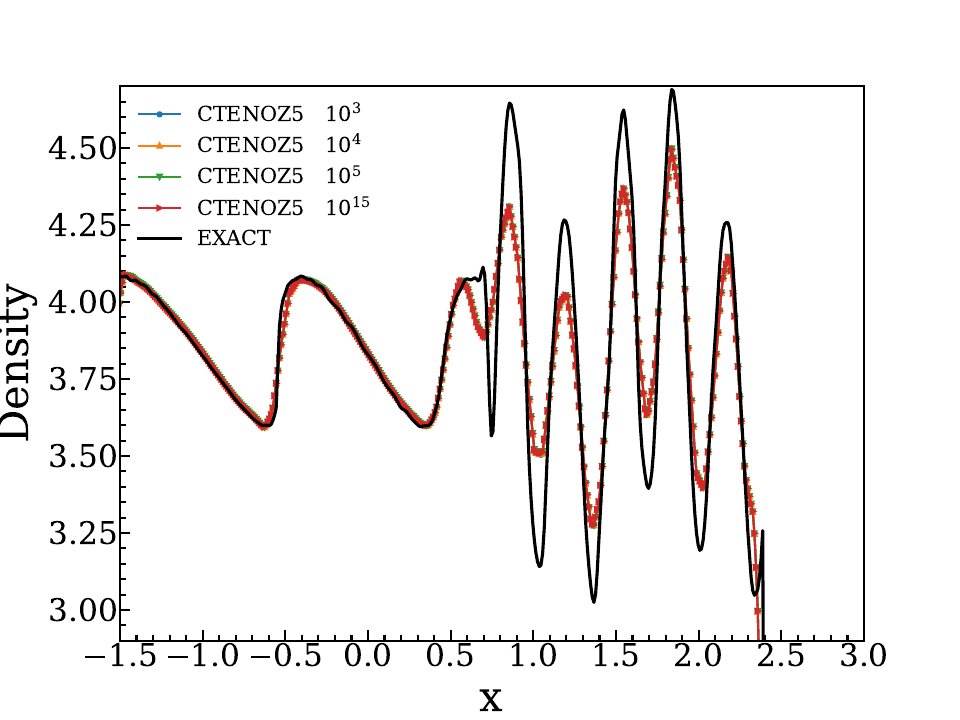}}
     \subfigure[]{
     \label{}
    \includegraphics[scale=0.4]{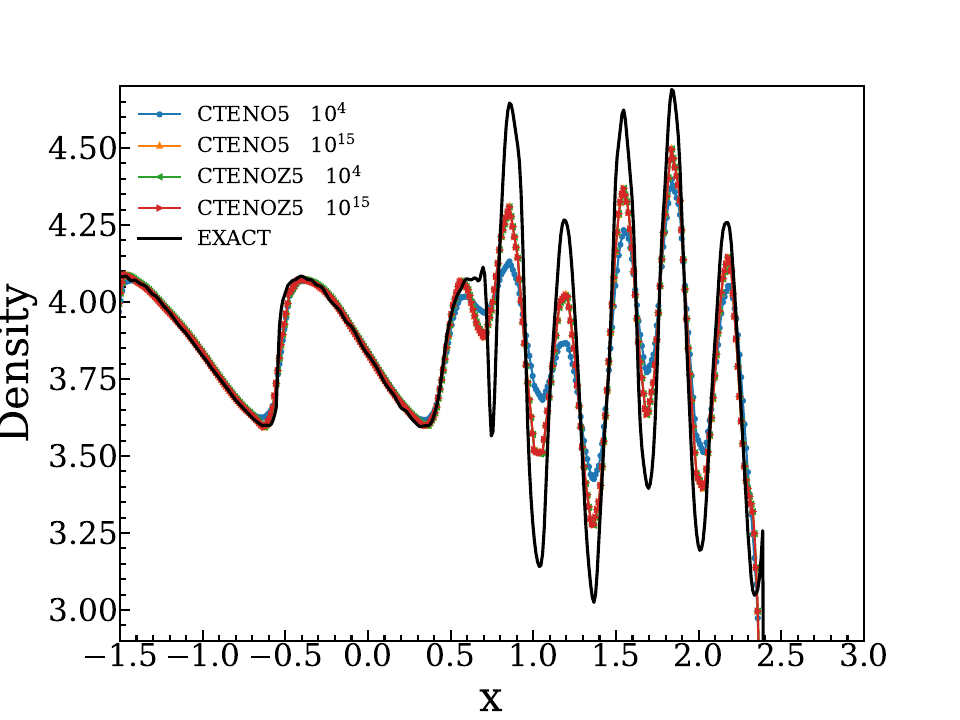}}
     \subfigure[]{
    \label{}
    \includegraphics[scale=0.4]{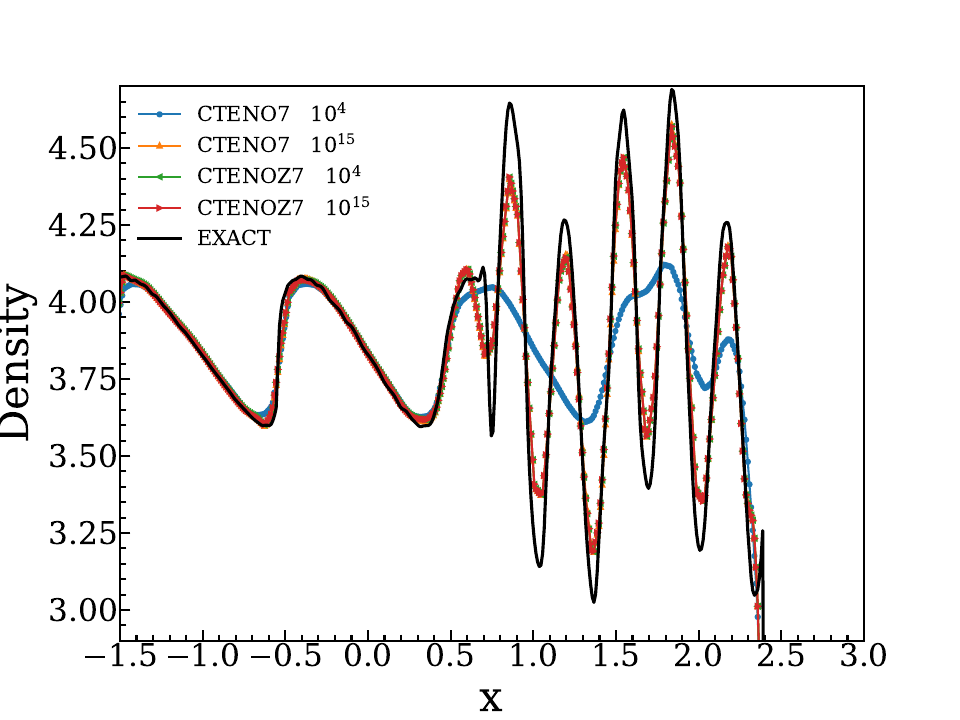}}
   
	\caption{shock-density wave interaction: 5th and 7th order solutions from  CTENO and CTENOZ with different \(\lambda_{1} ^{'}\). Density profiles are compared with reference solution.}
	\label{fig5}
\end{figure*}

As noted by Tsoutsanis and Dumbser \cite{8}, CWENO schemes offer the flexibility to adjust the central stencil linear weight $\lambda_{1}^{'}$. To comprehensively compare the performance of 5th order CTENO and CTENOZ schemes across various $\lambda_{1}^{'}$ values, density profiles for CTENO and CTENOZ schemes with values of $10^{3}$, $10^{4}$, $10^{5}$, and $10^{15}$ are depicted in \autoref{fig5} (a) and (b). Notably, in CTENO schemes, higher $\lambda_{1}^{'}$ values exhibit improved resolution for high-wavenumber fluctuations, echoing findings from numerical experiments conducted on CWENO schemes \cite{28}. On the other hand, CTENOZ schemes yield almost identical results across different $\lambda_{1}^{'}$ values, showcasing insensitivity to the choice of central stencil linear weight. This robustness of CTENOZ scheme can be attributed to the scaling of smoothness indicators derived from polynomials of varying orders, which helps to reach high order accuracy with different $\lambda_{1}^{'}$. While larger $\lambda_{1}^{'}$ values in CWENO and CTENO schemes enhance resolution of high-wavenumber fluctuations, they come at the cost of robustness, as demonstrated in additional numerical experiments on CWENO schemes under scenarios involving extremely strong discontinuities \cite{28}.

Concluded from the systematic testing of $\lambda_{1}^{'}$, it has been observed that CTENO scheme is significantly influenced by $\lambda_{1}^{'}$ and only attains theoretical accuracy at very large values of $\lambda_{1}^{'}$. Conversely, CTENOZ scheme is less sensitive to $\lambda_{1}^{'}$ and can achieve 5th-order accuracy for all mesh resolution and 7th order for coarse grid at $\lambda_{1} ^{'}=10^{4}$. Furthermore, CTENOZ scheme exhibits better robustness in different $\lambda_{1}^{'}$. Therefore, in the practical simulation, it is essential to balance both accuracy and robustness when selecting the appropriate $\lambda_{1}^{'}$. In the subsequent tests, we employed a value of $\lambda_{1} ^{'}=10^{15}$ for several test cases to attain the highest 7th-order accuracy for both CTENO and CTENOZ scheme. Additionally, in certain tests, we compared the performance of different schemes under $\lambda_{1} ^{'}=10^{4}$, ensuring that CTENOZ scheme can achieve either 5th or 7th-order of accuracy.

\subsection{Shock-tube problem}
\label{}

The 1D shock-tube problem has been extensively employed for evaluating implementation of numerical methods \cite{39}. The provided initial conditions are detailed below:

\begin {equation}
(\rho ,u,p)=\begin{cases}
	(1,0,1), \ -0.5\le x< 0 \\
	(0.125,0,0.1). \  0\le x\le 0.5
\end{cases}
\end {equation}

The computational domain spans from \([-0.5, 0.5]\times [0, 0.2]\). In this case, unstructured elements are utilized with an approximate resolution of \(h \approx  1/68\). The high-order reconstruction uses characteristic variables. The simulation is conducted on an unstructured mesh comprising of 2112 uniformly triangular cells. The ultimate computation time is set at \(t = 0.2\). $\lambda_{1} ^{'}=10^{15}$ is applied in this test for 7th order of accuracy.

\begin{figure*}[h]
	\centering
	
	\subfigure[]{
		\label{}
		\includegraphics[scale=0.4]{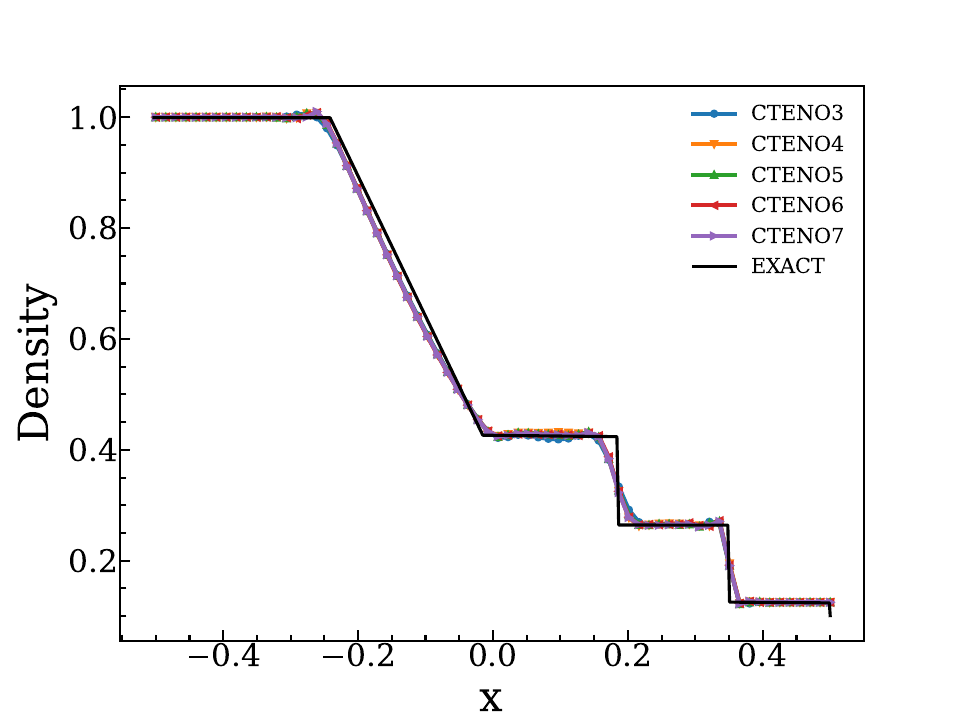}}
	\subfigure[]{
		\label{}
		\includegraphics[scale=0.4]{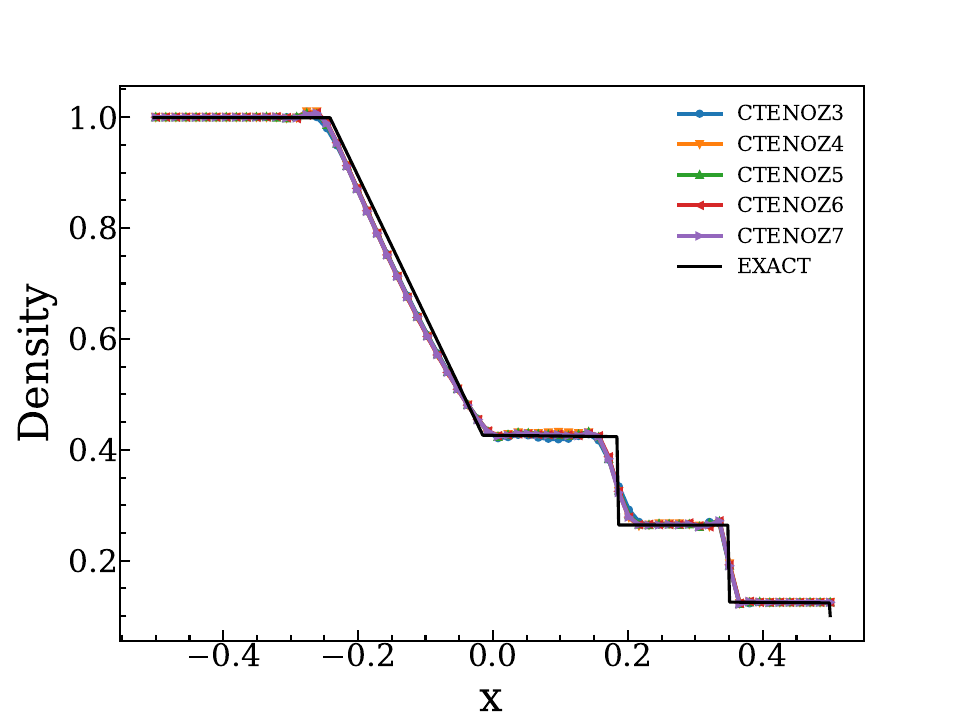}}

	\caption{shock-tube problem: solutions from CTENO and CTENOZ schemes of different orders. Density profiles are compared with precise solution.}
	\label{fig2}
\end{figure*}

\begin{figure*}[h]
	\centering
	
	\subfigure[]{
		\label{}
		\includegraphics[scale=0.4]{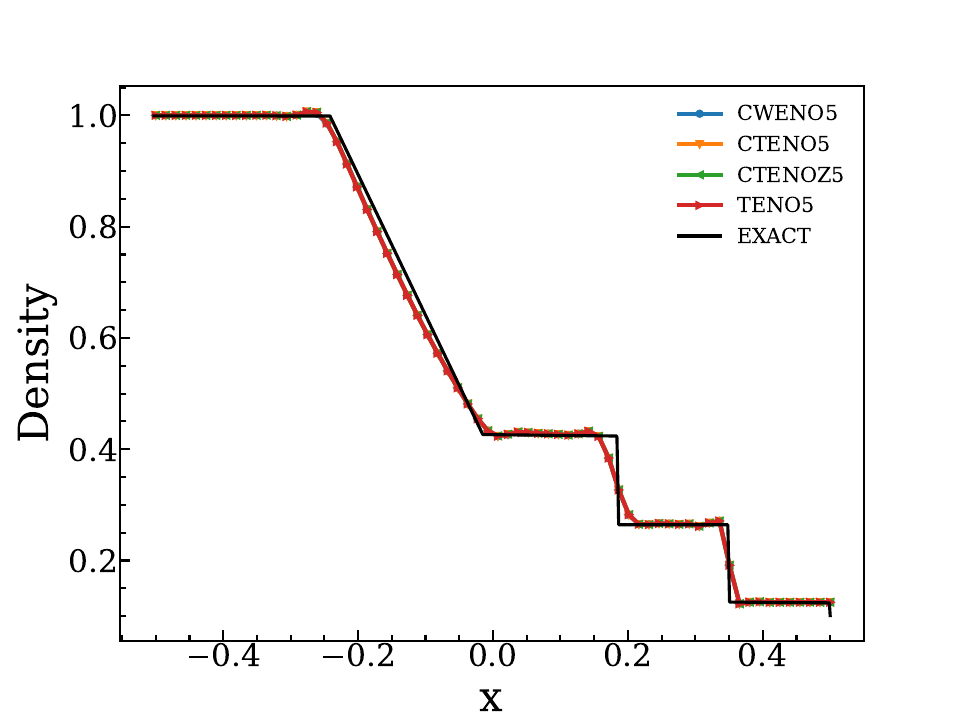}}
	\subfigure[]{
		\label{}
		\includegraphics[scale=0.4]{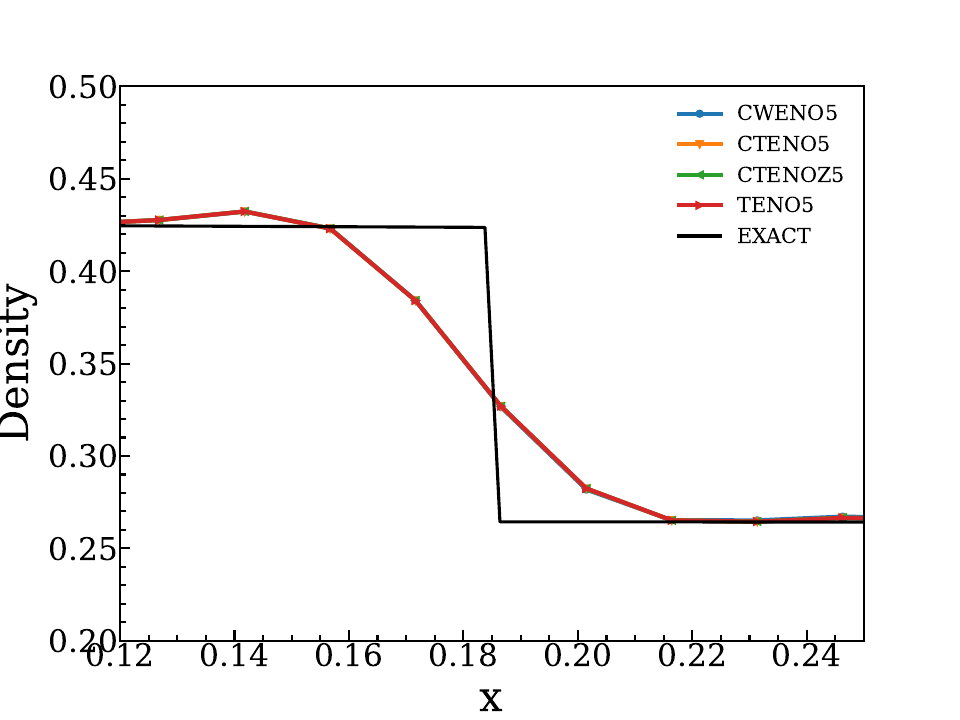}}

	\caption{shock-tube problem: solutions from CWENO,TENO,CTENO and CTENOZ schemes with 5th-order. Density profiles are compared with precise solution.}
	\label{fig3}
\end{figure*}

The density profiles obtained from CTENO and CTENOZ of varying orders are illustrated in \autoref{fig2}, effectively capturing the discontinuities. The precise solution to the Sod shock tube was obtained from Riemann problem. \autoref{fig3} showcases density distributions using different schemes, with all schemes showing good agreement with the exact solution, accurately capturing discontinuities and shock waves. Although the considered schemes in this test adapt lower-order directional stencils when encountering discontinuities, the order of directional stencils is more than first-order accuracy, rendering it more susceptible to oscillations. In subsequent test scenarios, both CTENO and CTENOZ schemes demonstrate robust non-oscillatory behavior.

\subsection{2D Double Mach Reflection of a Strong Shock}
\label{}

This classical 2D double Mach reflection of a strong shock is referenced from \cite{41}. It is commonly employed for assessing the properties of dissipation and capability to capture discontinuous. The initialization of computational domain is:

\begin {equation}
\begin{array}{c}
(\rho ,u,v,p)=
\begin{cases}
  (1.4,0,0,1), \ \text{ if } y< 1.732(x-0.1667) \\
  (8,7.145,-4.125,116.8333). \ \text{ otherwise } 
\end{cases}
\end{array}
 \end {equation}

The computational domain spans from \([0, 4]\times[0, 1]\) and consists of 369604 uniformly distributed triangular cells, with each cell approximately \(h \approx 1/200\). The simulation runs until \(t = 0.2\). A shock wave with Mach number of \(10\) originates from \(x = 0.1667\) and has an incident angle of \(60\) degrees relative to \(x\)-axis, travels towards right. The initial state of post-shock is enforced from \(x = 0\) to \(x = 0.1667\), while wall boundary condition is applied from \(x = 0.1667\) to \(x = 4\) at bottom. The fluid are defined at the top to accurately capture the behavior of shockwave. 

\begin{figure*}[h]
	\centering
 
    \subfigure[CWENO5]{
    \label{}
    \includegraphics[scale=0.19]{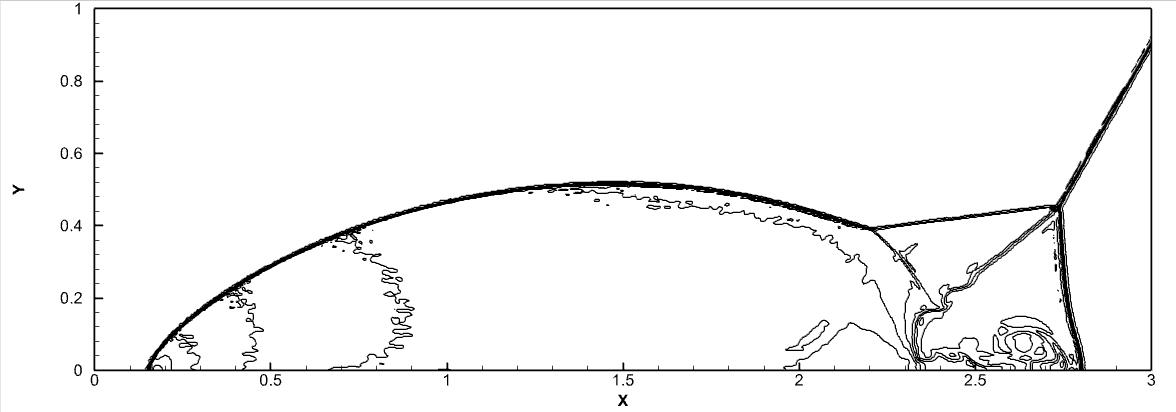}}
    \subfigure[TENO5]{
    \label{}
    \includegraphics[scale=0.19]{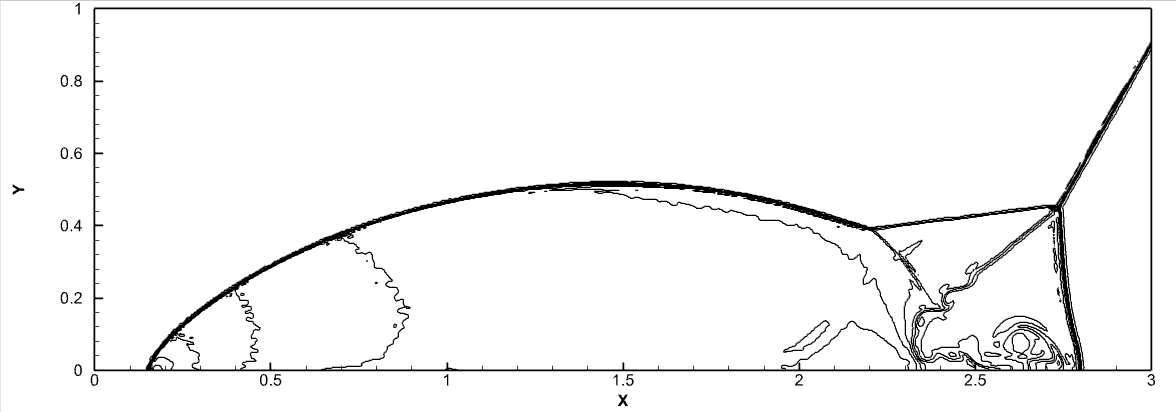}}
    \subfigure[CTENO5]{
    \label{}
    \includegraphics[scale=0.19]{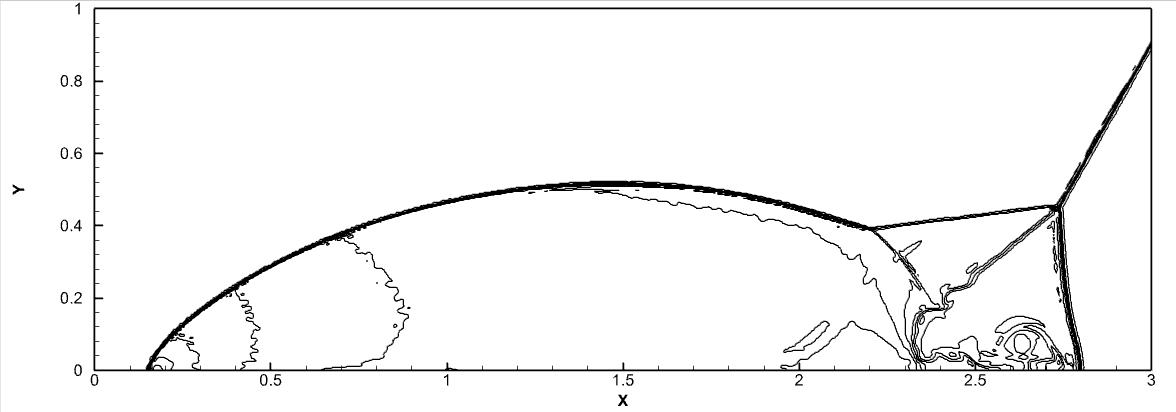}}
    \subfigure[CTENOZ5]{
    \label{}
    \includegraphics[scale=0.19]{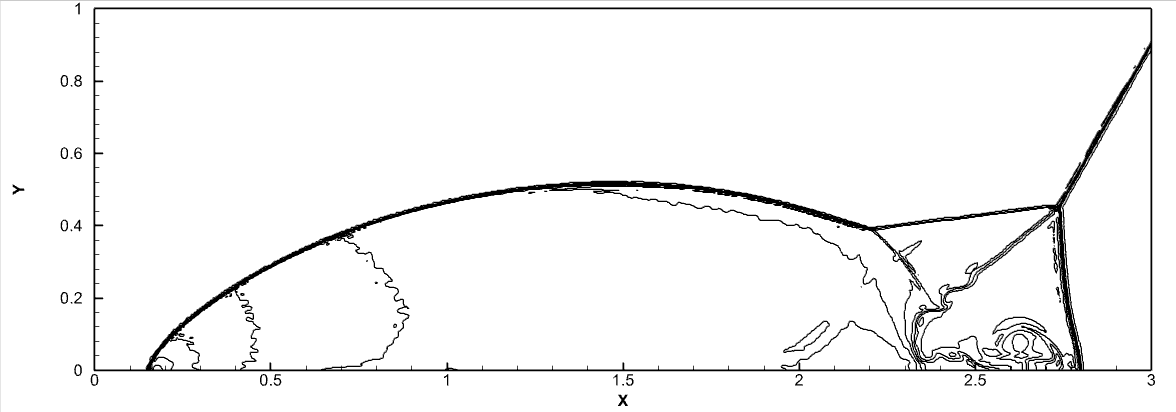}}
    \subfigure[CWENO7]{
    \label{}
    \includegraphics[scale=0.19]{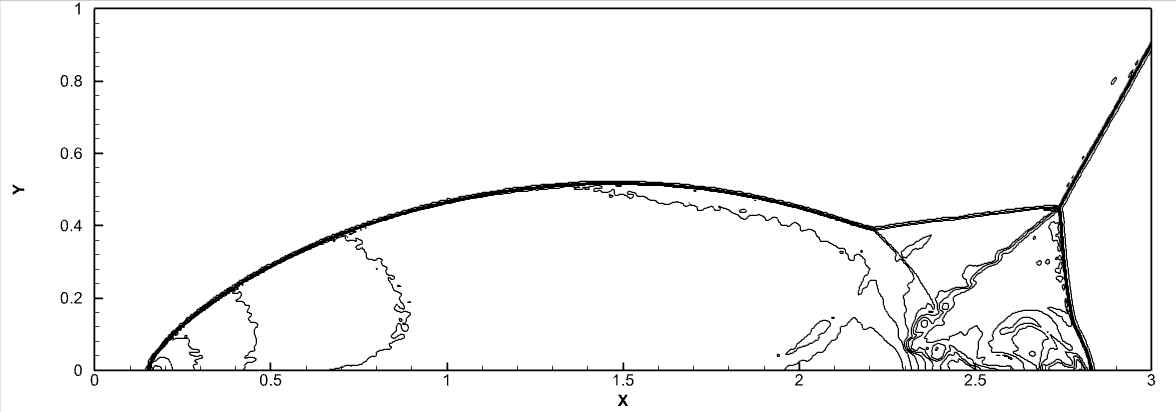}}
    \subfigure[TENO7]{
    \label{}
    \includegraphics[scale=0.19]{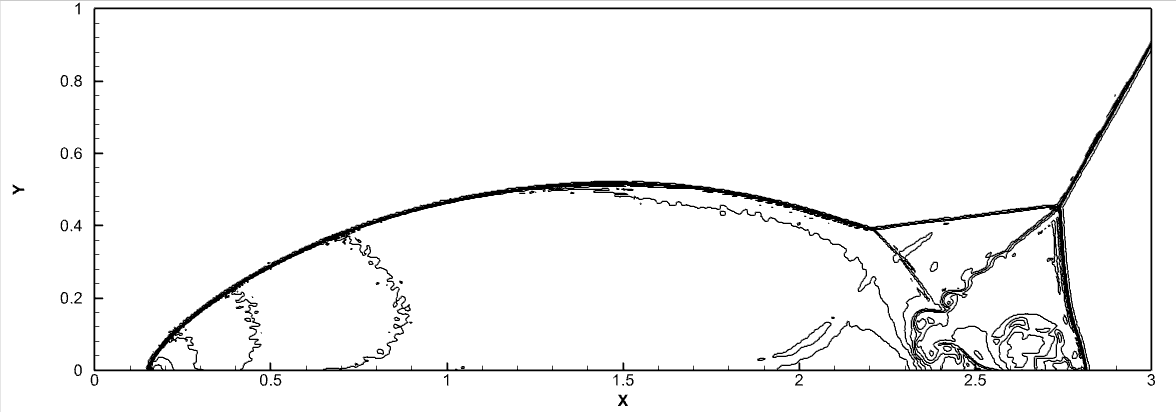}}
     \subfigure[CTENO7]{
    \label{}
    \includegraphics[scale=0.19]{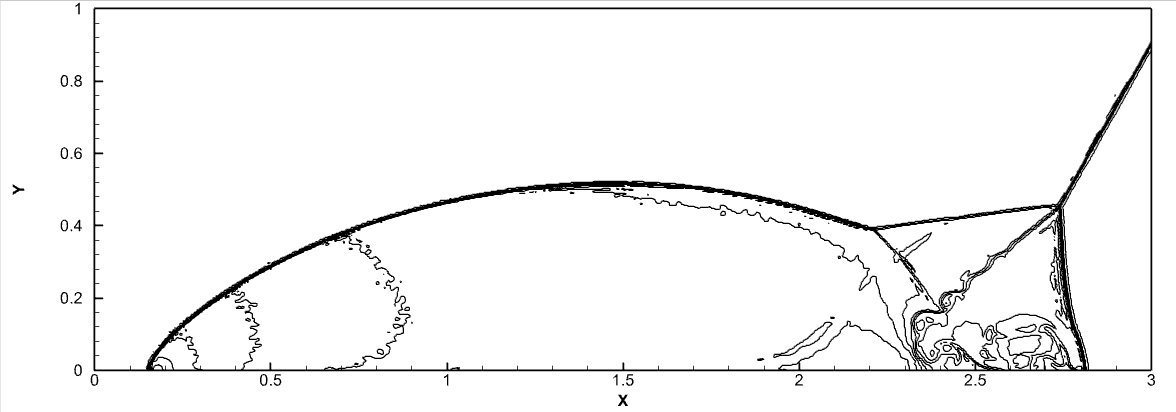}}
    \subfigure[CTENOZ7]{
    \label{}
    \includegraphics[scale=0.19]{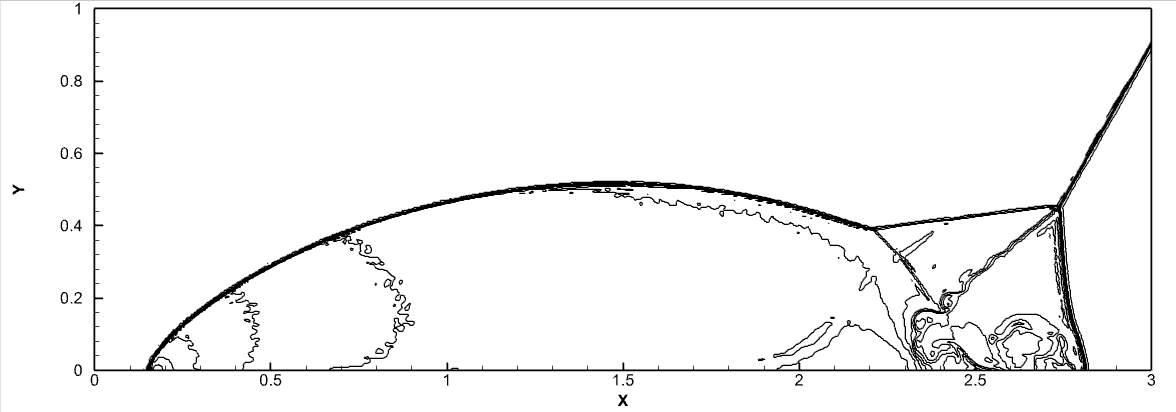}}

	\caption{2D double Mach reflection of a strong shock: results from (a) CWENO5, (b) TENO5,(c)CTENO5,(d) CTENOZ5,(e) CWENO7 ,(f)TENO7,(g)CTENO7 and (h)CTENOZ7. The figures contain density contours ranging from 1.5 to 21.5.}
	\label{fig6}
\end{figure*}

\begin{figure*}[h]
	\centering
	
	\subfigure[CWENO5]{
		\label{}
		\includegraphics[scale=0.09]{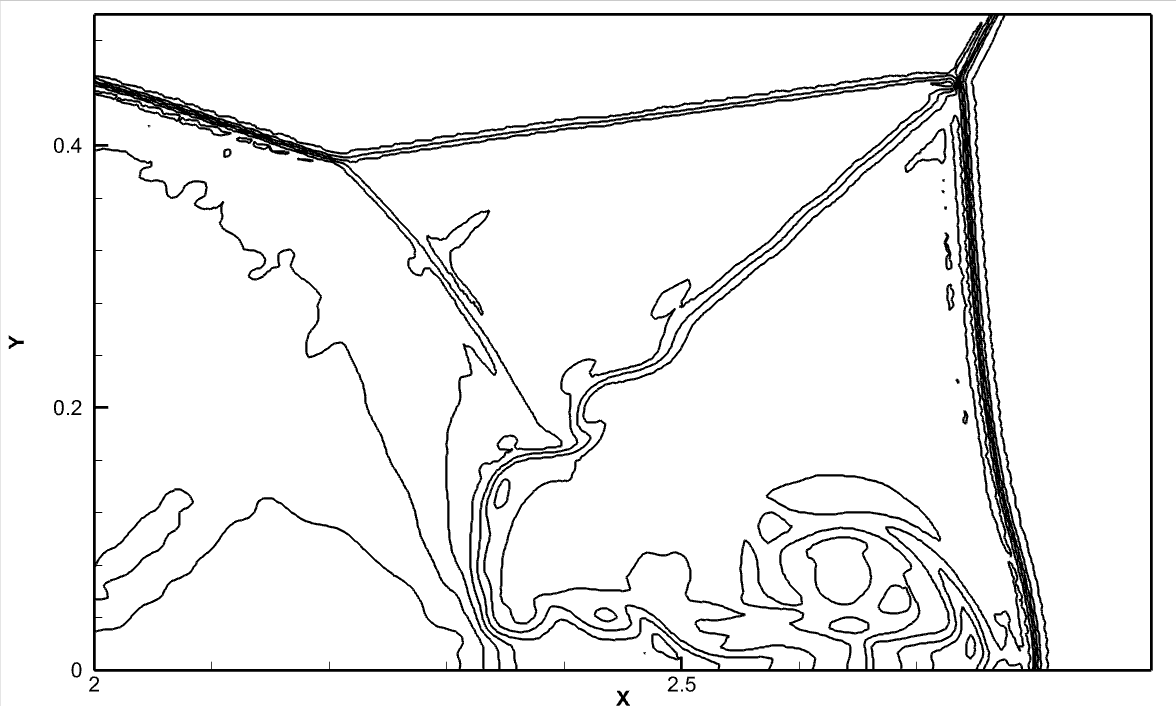}}
	\subfigure[TENO5]{
		\label{}
		\includegraphics[scale=0.09]{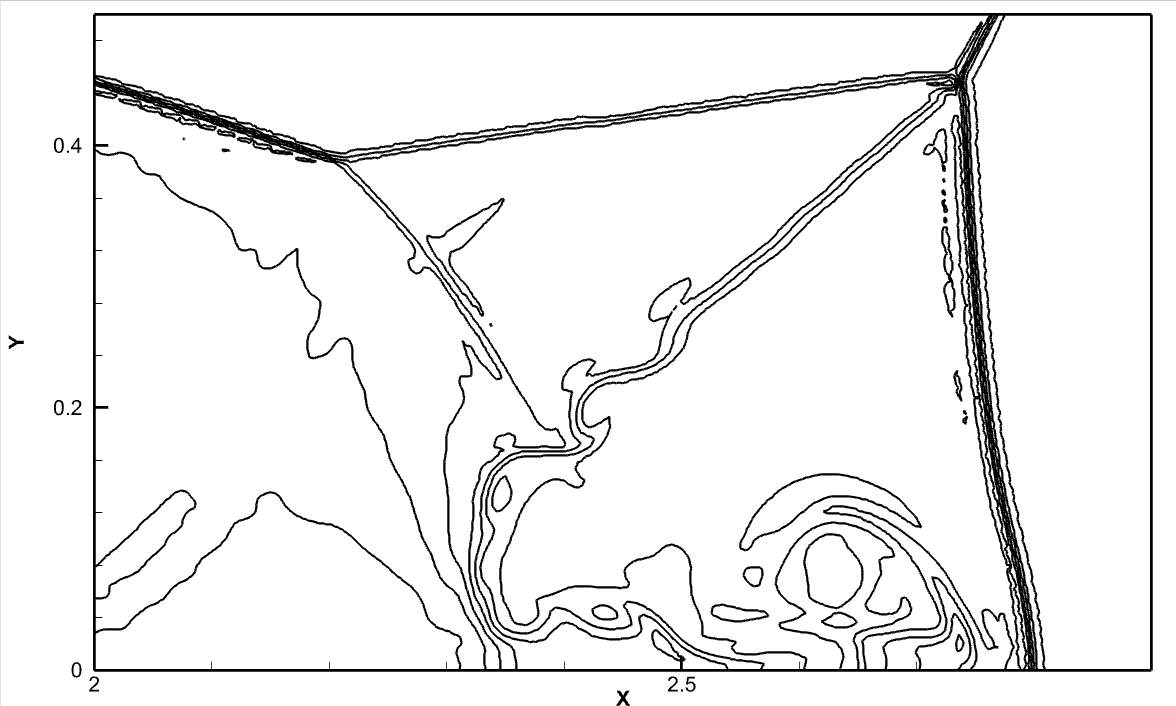}}
	\subfigure[CTENO5]{
		\label{}
		\includegraphics[scale=0.09]{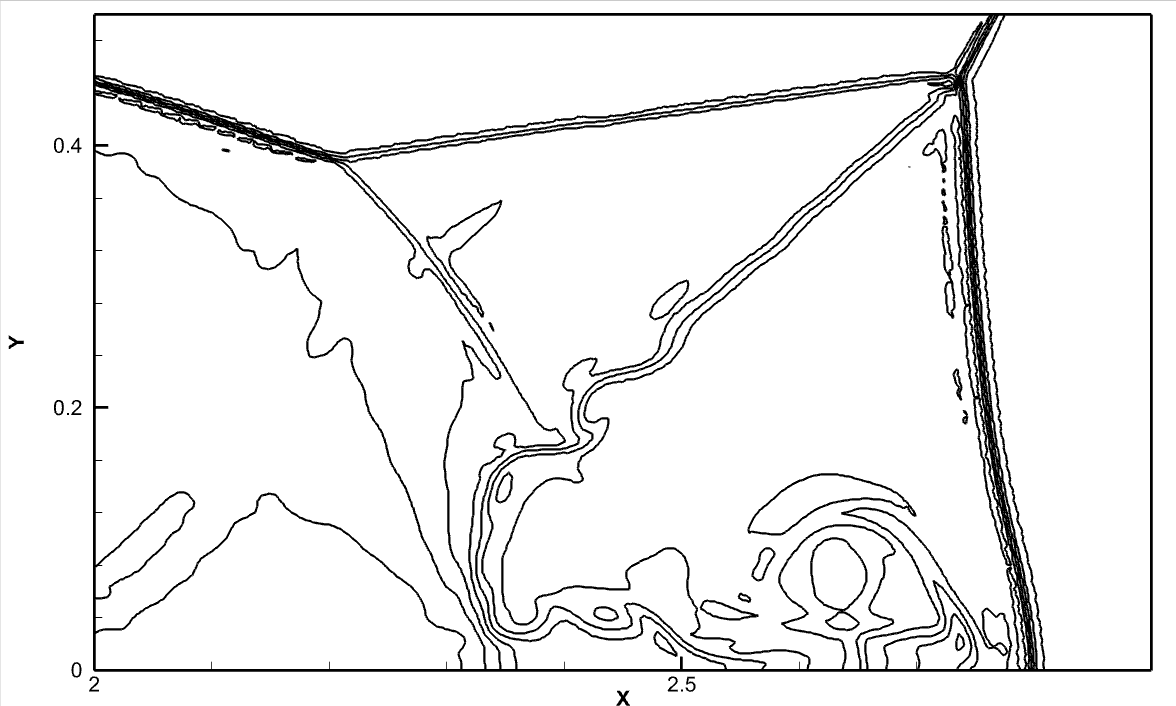}}
	\subfigure[CTENOZ5]{
		\label{}
		\includegraphics[scale=0.09]{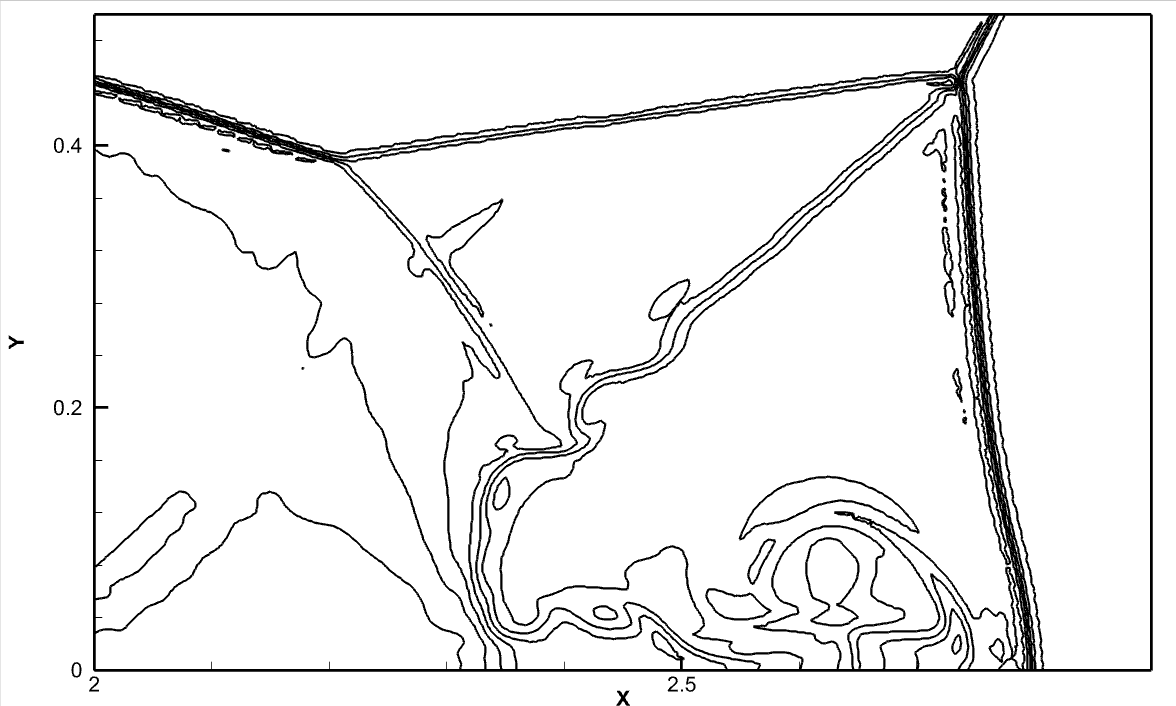}}
	\subfigure[CWENO7]{
		\label{}
		\includegraphics[scale=0.09]{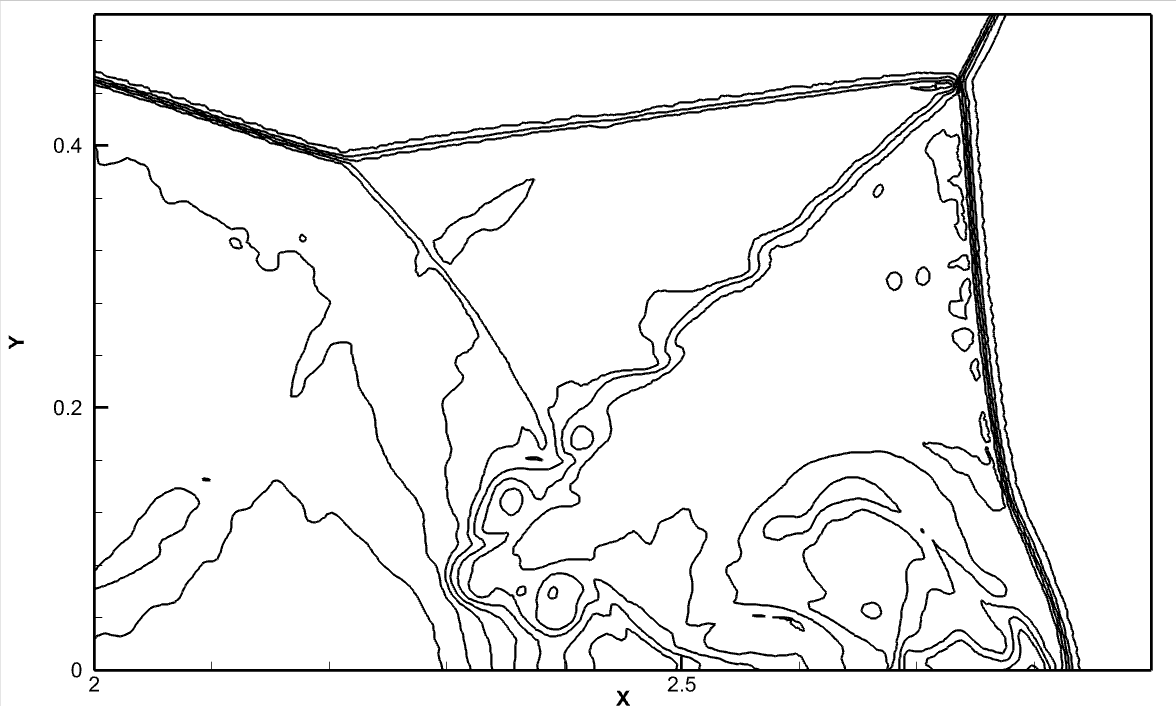}}
	\subfigure[TENO7]{
		\label{}
		\includegraphics[scale=0.09]{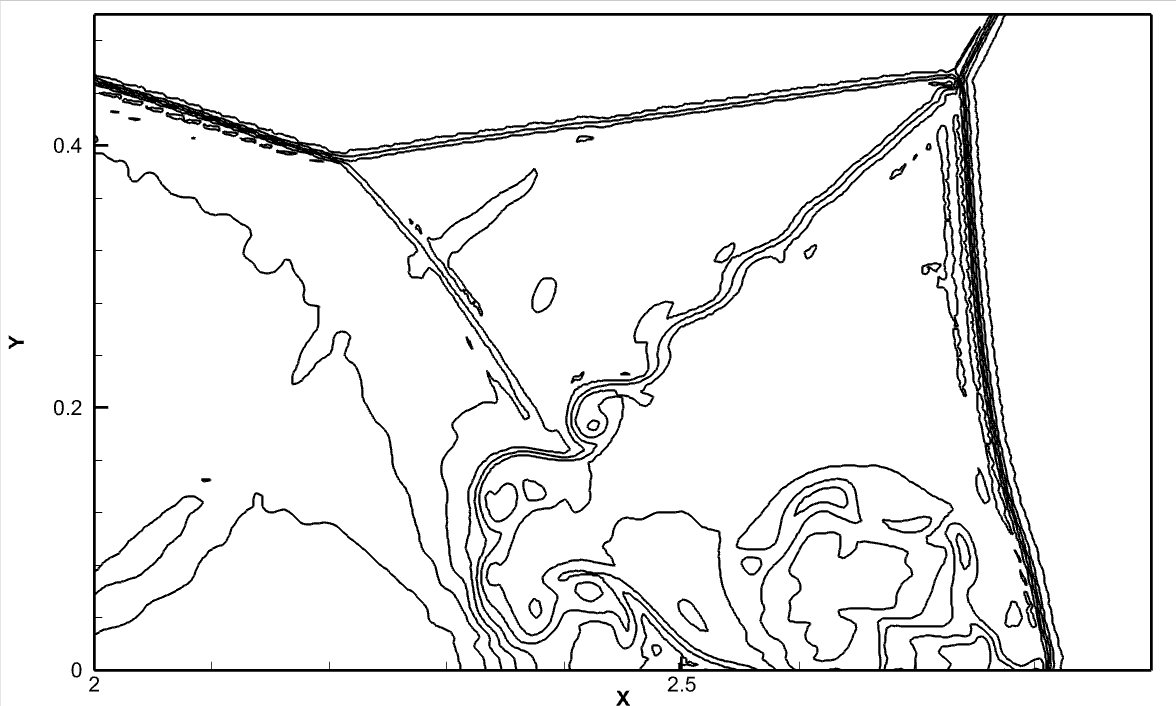}}
	\subfigure[CTENO7]{
		\label{}
		\includegraphics[scale=0.09]{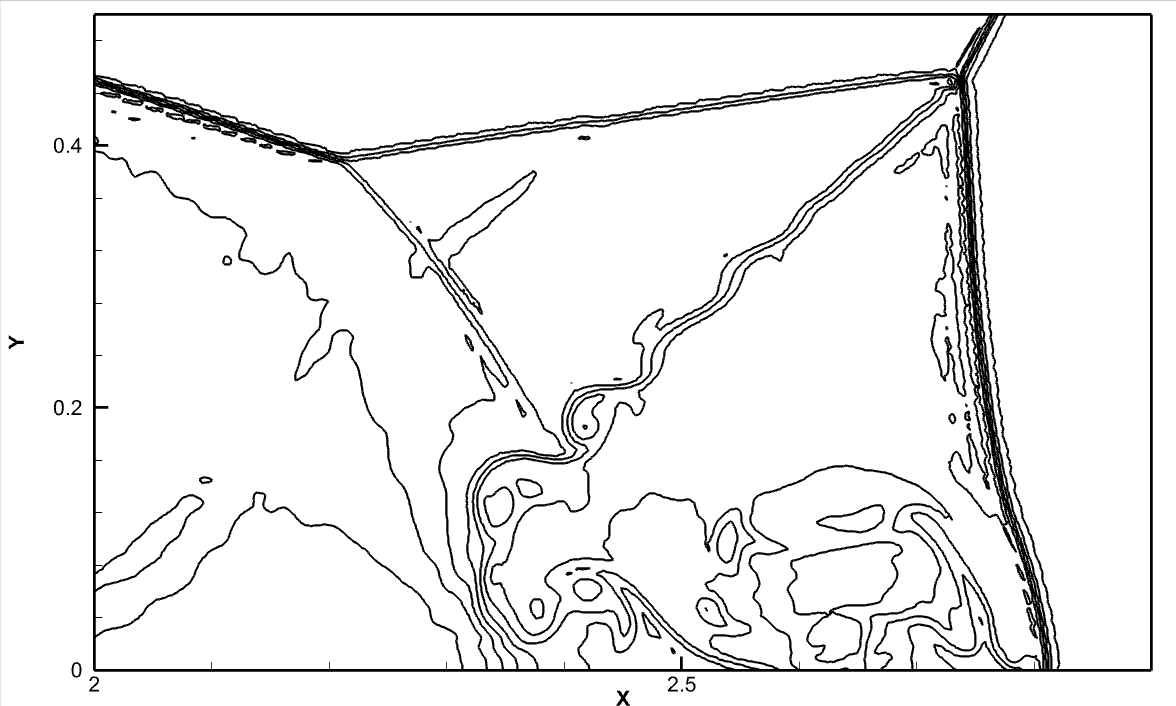}}
	\subfigure[CTENOZ7]{
		\label{}
		\includegraphics[scale=0.09]{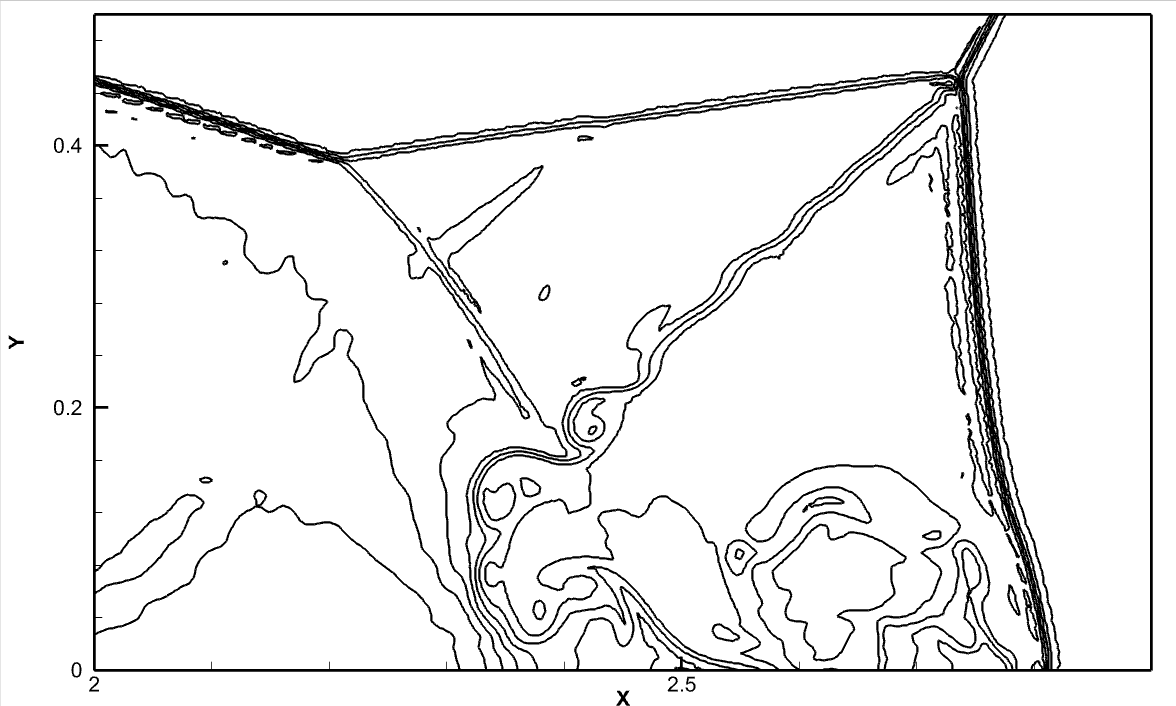}}

	\caption{2D double Mach reflection of a strong shock: results from (a) CWENO5, (b) TENO5,(c)CTENO5,(d) CTENOZ5,(e) CWENO7 ,(f)TENO7,(g)CTENO7 and (h)CTENOZ7. A close-up view of the Mach stem. The figures contain density contours ranging from 1.5 to 21.5.}
	\label{fig7}
\end{figure*}

\autoref{fig6} and \autoref{fig7} illustrates the numerical results for 5th and 7th orders of CTENO and CTENOZ schemes and their respective CWENO and TENO schemes with $\lambda_{1} ^{'}=10^{15}$ for achieving 7th order of accuracy. As shown in \autoref{fig6} and \autoref{fig7}, higher-order CTENO and CTENOZ schemes can accurately capture small-scale features with greater precision and reduced numerical dissipation. The density distributions reveals the absence of any erroneous numerical oscillations near the discontinuities. It is important to highlight that CWENO, TENO, CTENO, and CTENOZ schemes demonstrate robustness outperforming the WENO scheme \cite{42}. The region surrounding the double Mach stems showed significant variations in the results, with CTENO and CTENOZ demonstrating low dissipation and facilitating identification of small-scale vortex structures.

\subsubsection{run time}
\label{}

The computational times presented in 2D double Mach reflection were derived from identical hardware (and compilation settings) and standardized against a reference setup on the same hardware. The unstructured meshes consist of 369604 and 714138 uniform triangular elements are used for run time test. This normalization allows for a fair comparison of algorithm performance under identical hardware conditions. The benefits of the CWENO schemes, in contrast to the WENO scheme, are not solely limited to reduced computational time but also exhibit significantly improved robustness \cite{8,9}. In this particular test, the CWENO ,TENO, CTENO, and CTENOZ scheme require approximately the same amount of time with the same spatial order, as shown in \autoref{tab1}.

\begin{table}[h]
\centering
\caption{Normalised computational times (with respect to CWENO 5th order with 369604 uniform triangular elements) for 2D Double Mach Reflection test problem.}
\begin{tabular}{ccccc}
\toprule
Elements & CWENO5 & TENO5 & CTENO5 & CTENOZ5 \\
\midrule
369604 & 1.000 & 1.036 & 1.033 & 1.000 \\
714138 & 2.798 & 2.870 & 2.795 & 2.811 \\
\bottomrule
                 
\end{tabular}
\label{tab1}
\end{table}

\subsubsection{Parallel scalability}
\label{}

The performance of parallelization for existing methods and computer code was evaluated by implementing the 2D double Mach reflection flow case with the finest mesh consisting of 714138 triangular elements. Grid partitioning was performed using the METIS software \cite{60}. The numerical methods used in this study were incorporated into Fortran 90 code, which employed the Message Passing Interface (MPI) for parallel communications. The performance tests were conducted at the Beijing Super Cloud Computing Center (BSCC) facility. The numerical experiments were carried out on the AMD EPYC 7542 CPU model. The maximum computational time for 2000 iterations was recorded for all test cases.

The original code implementation shows that the calculation of WENO weights takes up the most time, followed by least-square reconstruction and extrapolation of flow variables and gradients at cell interfaces. When considering communication expenses, the most expensive aspect lies in communication of the reconstructed solutions and their gradients for the Gaussian quadrature points between inter-processor boundaries, rather than communication of the solutions for the stencil elements across inter-processor boundaries. Therefore, the calculation of WENO weights has been recognized as having the capability to greatly improve the performance of the CFD code \cite{61}. This research focuses on examining the weights calculation of CTENO and CTENOZ schemes in order to assess their parallel performance.

\begin{figure*}[h]
	\centering
    \includegraphics[scale=0.5]{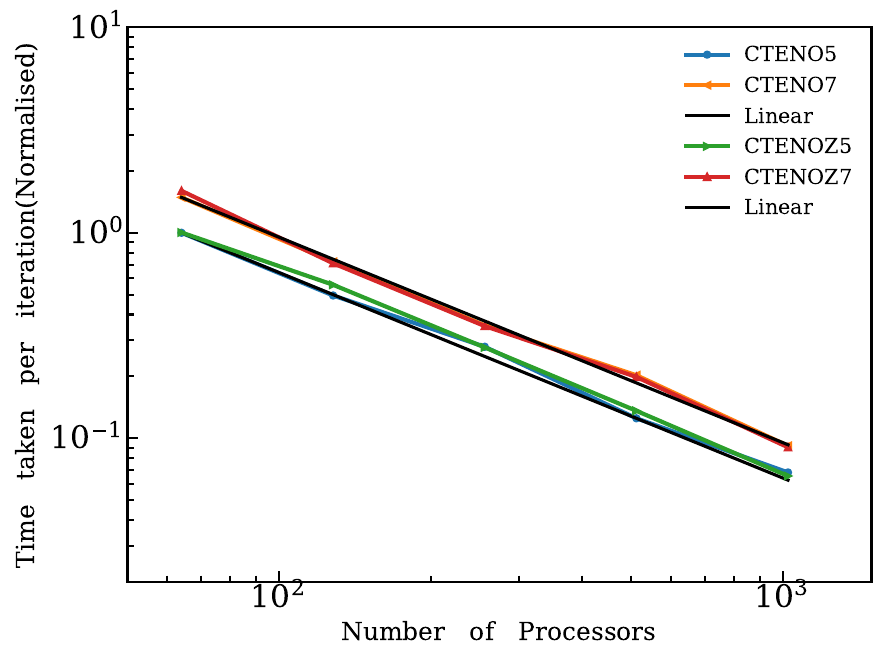}
	\caption{Parallel scalability of CTENO5, CTENO7, CTENOZ5 and CTENOZ7 numerical schemes for the 2D double Mach reflection.}
	\label{scaling}
\end{figure*}

The reference time for normalizing the CPU results of the numerical simulations was calculated based on the time required for one time-iteration of the CTENO-5th order scheme on 64 processors. To assess the scalability of the methods and computer code, experiments were conducted with the number of processors ranging from 64 to 1024. As the numerical accuracy order increases, the computational time increases at a faster rate compared to the communication time, indicating that higher-order numerical schemes exhibit better scalability than lower-order schemes. This is illustrated in \autoref{scaling}.

The parallelization of the unstructured schemes involves interprocessor communication for exchanging two types of data: (i) the cell-centered values of halo elements on the stencil, and (ii) the reconstructed, boundary extrapolated values of Gaussian quadrature points on the direct-side halo neighboring elements. The second communication requirement is commonly found in high-order schemes. The amount of data needed for (i) is much smaller compared to (ii) because the latter involves the boundary extrapolated values of the solution and solution gradients at each surface Gaussian quadrature point. For the CTENO-5th order scheme, the wall-clock time per iteration is 1.7s on 64 processors, and for the CTENO-7th order scheme, it is approximately 1.5 times higher. When the number of processors increases from 64 to 128, 256, 512, 1024, the iteration time decreases, resulting in an approximately 2.0 times speedup in practice \autoref{scaling}.

\subsection{Single-Material Triple Point Problem}
\label{}
We examine single-material triple point problem and adopt the configuration outlined in \cite{44}. The simulation region is \([0, 7] \times [0, 3]\) and initialization is specified as follows
\begin {equation}
\begin{array}{c}
(\rho ,u,v,p)=
\begin{cases}
  (1,0,0,1), \ \text{ if }0 \le  x\le 1 \\
  (0.125,0,0,0.1), \ \text{ if }1 < x\le7 \text{ and }1.5 < y\le 3 \\
  (1,0,0,0.1). \ \text{ if }1 < x\le 7\text{ and }0 \le  y\le1.5
\end{cases}
\end{array}
 \end {equation}

\begin{figure*}[h]
	\centering
 
   \subfigure[CWENO5]{
    \label{}
    \includegraphics[scale=0.17]{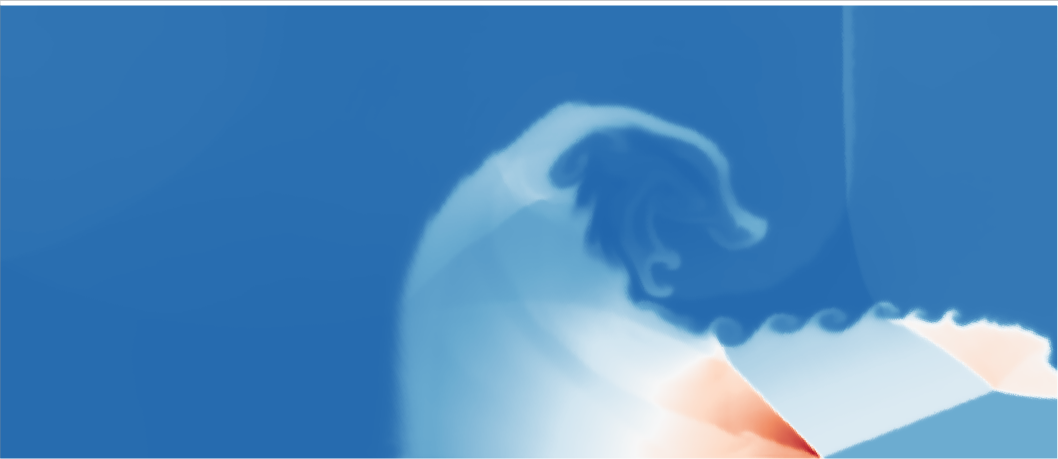}}
    \subfigure[TENO5]{
    \label{}
    \includegraphics[scale=0.17]{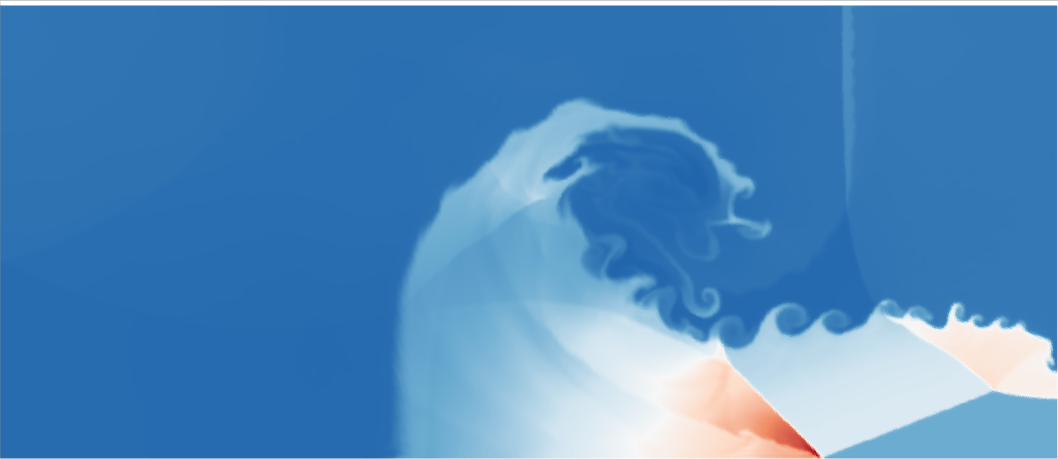}}
    \subfigure[CTENO5]{
    \label{}
    \includegraphics[scale=0.17]{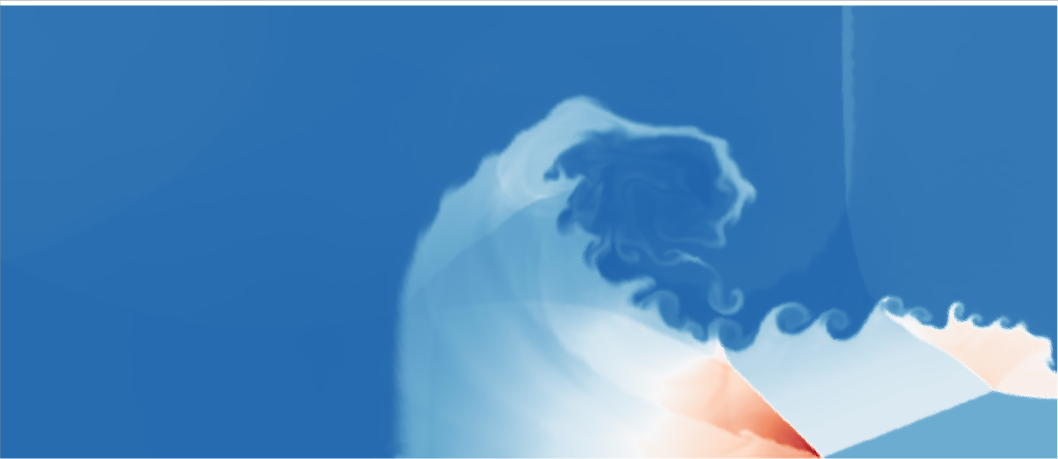}}
    \subfigure[CTENOZ5]{
    \label{}
    \includegraphics[scale=0.17]{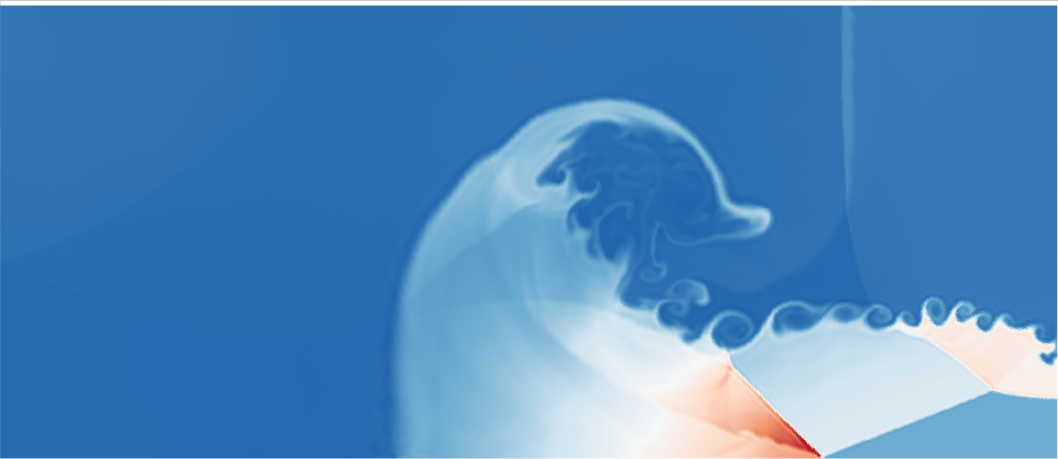}}
    \subfigure[CWENO7]{
    \label{}
    \includegraphics[scale=0.17]{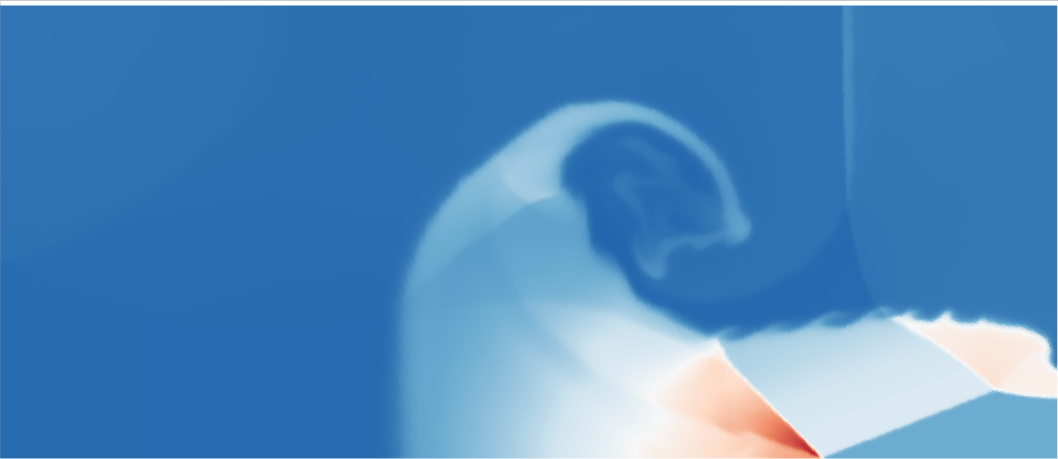}}
    \subfigure[TENO7]{
    \label{}
    \includegraphics[scale=0.17]{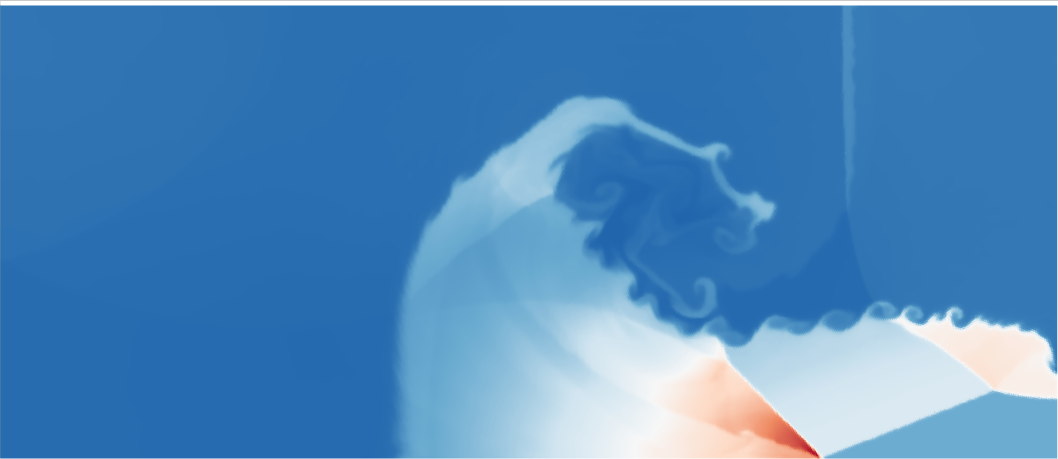}}
    \subfigure[CTENO7]{
    \label{}
    \includegraphics[scale=0.17]{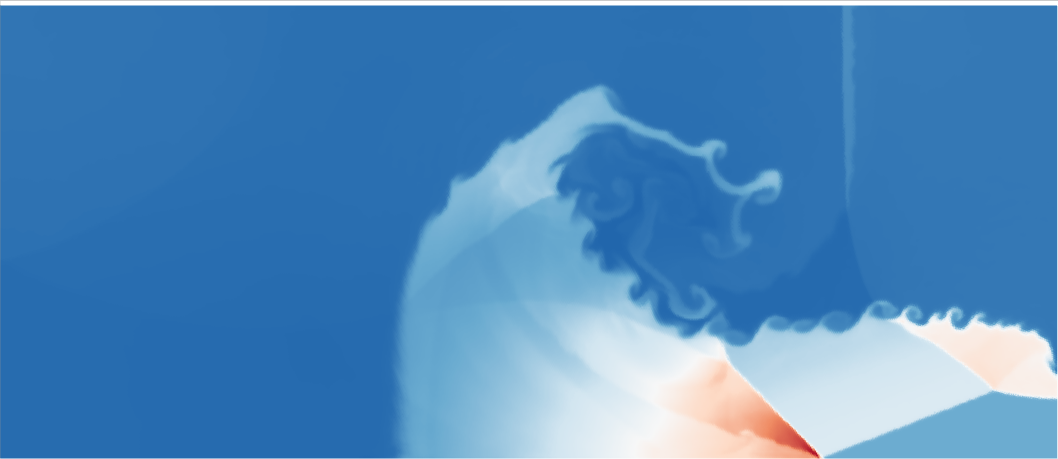}}
    \subfigure[CTENOZ7]{
    \label{}
    \includegraphics[scale=0.17]{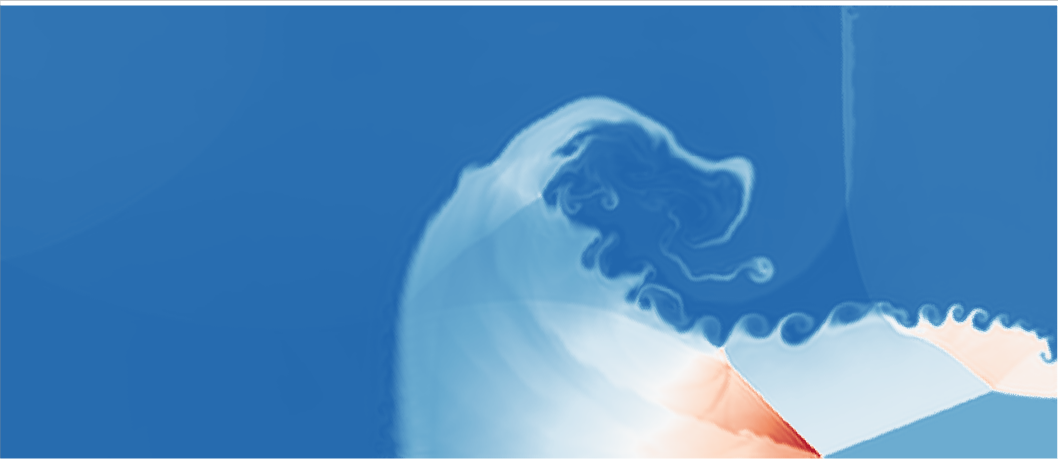}}
   
	\caption{Single-material triple point problem: results from the (a)CWENO5, (b)TENO5, (c)CTENO5,(d)CTENOZ5,(e)CWENO7, (f)TENO7, (g)CTENO7,and (h)CTENOZ7 schemes. The figures contain 25 density contours ranging from 0.2 to 5.0.}

	\label{fig8}
\end{figure*}

Wall conditions are applied at edges of the domain. Total time of simulation is \(t = 5\). The mesh comprises 214770 uniform triangular elements, with an estimated element size of \(h \approx 1/67\). $\lambda_{1} ^{'}=10^{4}$ is applied in this test for comparison with different schemes and ensuring 5th and 7th order of accuracy for CTENOZ. The solutions obtained from CWENO,TENO, CTENO and CTENOZ can be seen in \autoref{fig8}. The CTENO and CTENOZ schemes of high order achieve improved resolution of interfacial instability, incorporating smaller-scale structures. In comparison, the CTENOZ schemes outperform the corresponding other schemes, especially near the tips of the structures. The computational times needed for the CWENO ,TENO, CTENO, and CTENOZ schemes in this problem are also comparable to the time required for the same spatial order, as shown in \autoref{tab2}.

\begin{table}[h]
\centering
\caption{Normalised computational times (with respect to CWENO 5th order) for Single-Material Triple Point Problem.}
\begin{tabular}{ccccc}
\toprule
Order & CWENO & TENO & CTENO & CTENOZ \\
\midrule
5th & 1.000 & 1.024 & 1.043 & 1.081 \\
7th & 1.538 & 1.557 & 1.571 & 1.643 \\
\bottomrule
                 
\end{tabular}
\label{tab2}
\end{table}

\subsection{Kelvin-Helmholtz instability}
\label{}

This section examines 2D Kelvin-Helmholtz instability, which transforms a narrow shear layer into a complex arrangement of vortices. Such problems serve as a common benchmark to assess the precision and dissipation characteristics of numerical techniques in transforming a linear disturbance into a nonlinear state \cite{43}. The simulation domain corresponds to \([0, 1] \times [0, 1]\), and initial state is provided by:

\begin {equation}
\begin{array}{c}
(\rho ,u,v,p)=
\begin{cases}
  (2,-0.5,0.01sin(2\pi (x-0.5)),2.5), \ \text{ if }0.25 < y< 0.75 \\
  (1,0.5,0.01sin(2\pi (x-0.5)),2.5), \ \text{ otherwise } 
\end{cases}
\end{array}
 \end {equation}

where \(\gamma  = 1.4\) . Simulation concludes at a final time of \(t = 1\). The mesh consists of 577416 uniformly triangular cells with size of approximately \(h \approx 1/500\). $\lambda_{1} ^{'}=10^{4}$ is applied in this problem for comparison with different schemes and ensuring 5th order of accuracy for CTENOZ.

\begin{figure*}[h]
	\centering
 
   \subfigure[CWENO5]{
    \label{}
    \includegraphics[scale=0.15]{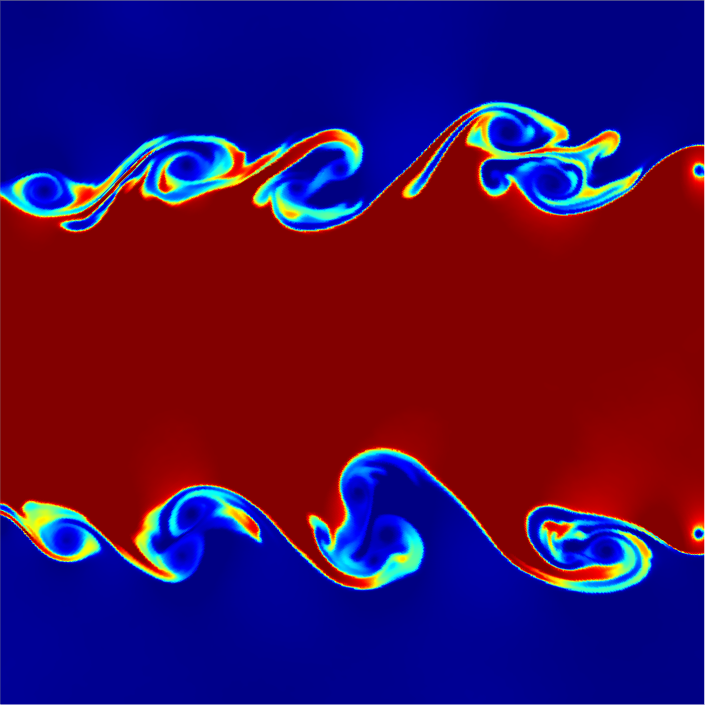}}
    \subfigure[TENO5]{
    \label{}
    \includegraphics[scale=0.15]{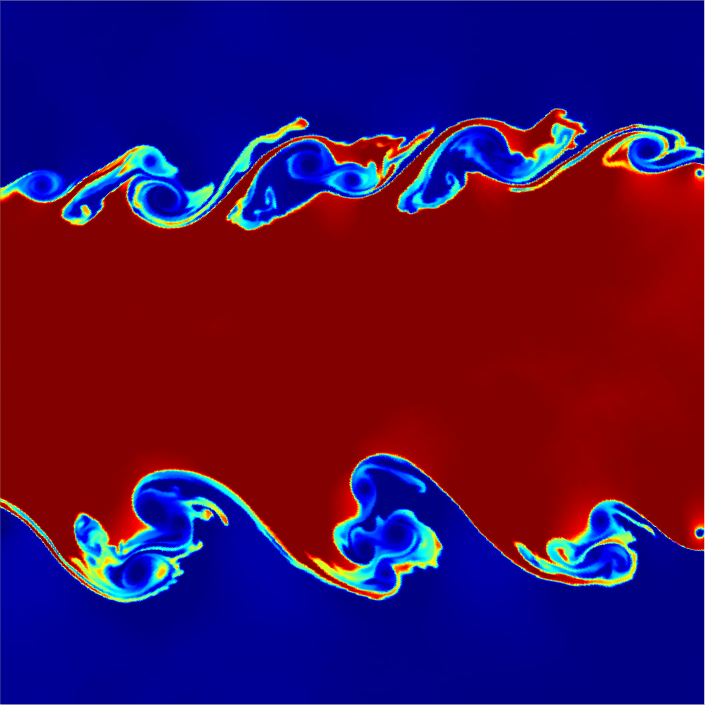}}
    \subfigure[CTENO5]{
    \label{}
    \includegraphics[scale=0.15]{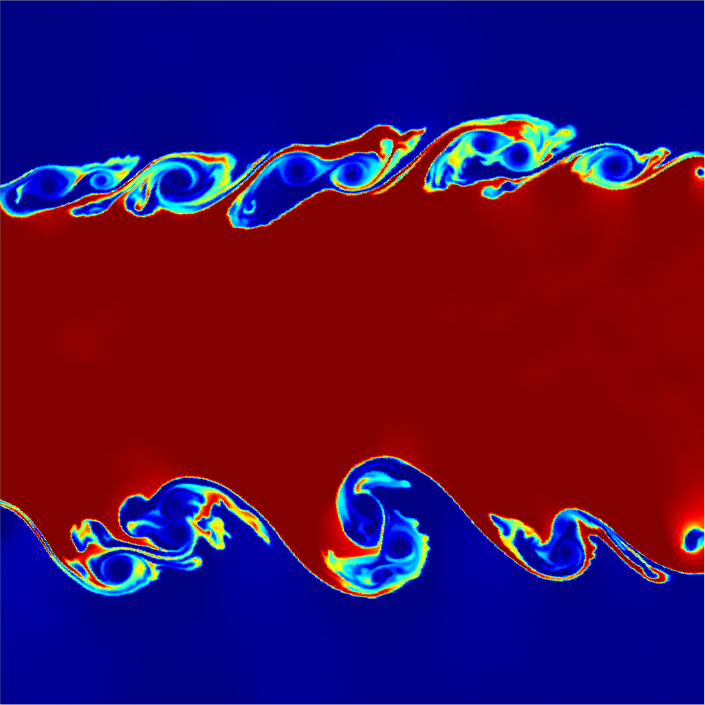}}
    \subfigure[CTENOZ5]{
    \label{}
    \includegraphics[scale=0.15]{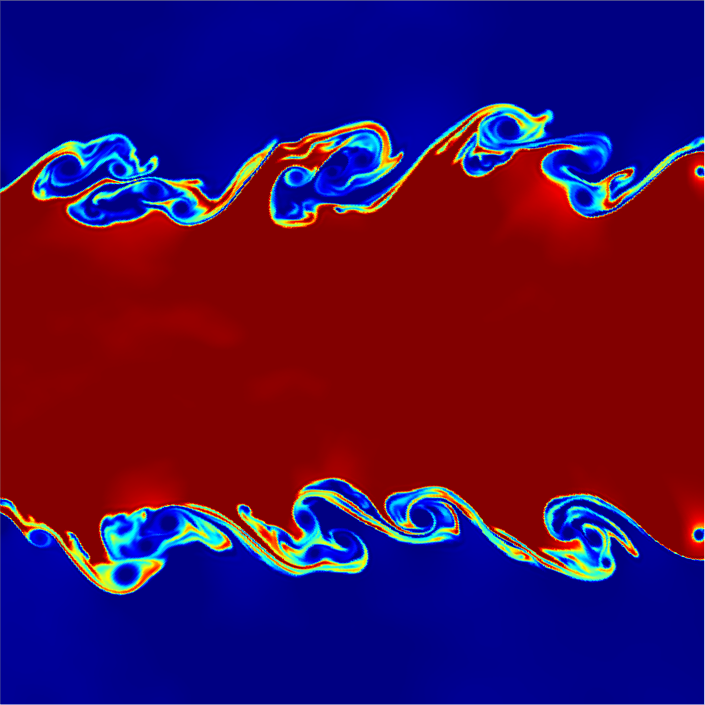}}

	 \caption{2D Kelvin-Helmholtz instability: density field from (a)CWENO5, (b)TENO5, (c)CTENO5,(d)CTENOZ5. The figures contain 13 density contours ranging from 0.9 to 2.1.}
	\label{fig9}
\end{figure*}

After comparing numerical results, the superior density distributions obtained with the CTENO and CTENOZ compared to CWENO and TENO are visible in \autoref{fig9}. 5th-order CTENO and CTENOZ scheme demonstrates enhanced accuracy, resulting in a significant resolution of the fine-scale vortical structures, as shown in \autoref{fig9}. It should be emphasized that the lack of a convergent solution arises from the interaction of instabilities, resulting in the emergence of new vortices and chaos during later stages. Additionally, discrepancies in mesh resolution and the choice of numerical methods further contribute to the distinctiveness of solutions achieved through different numerical methods.

\begin{figure*}[h]
	\centering

    \includegraphics[scale=0.4]{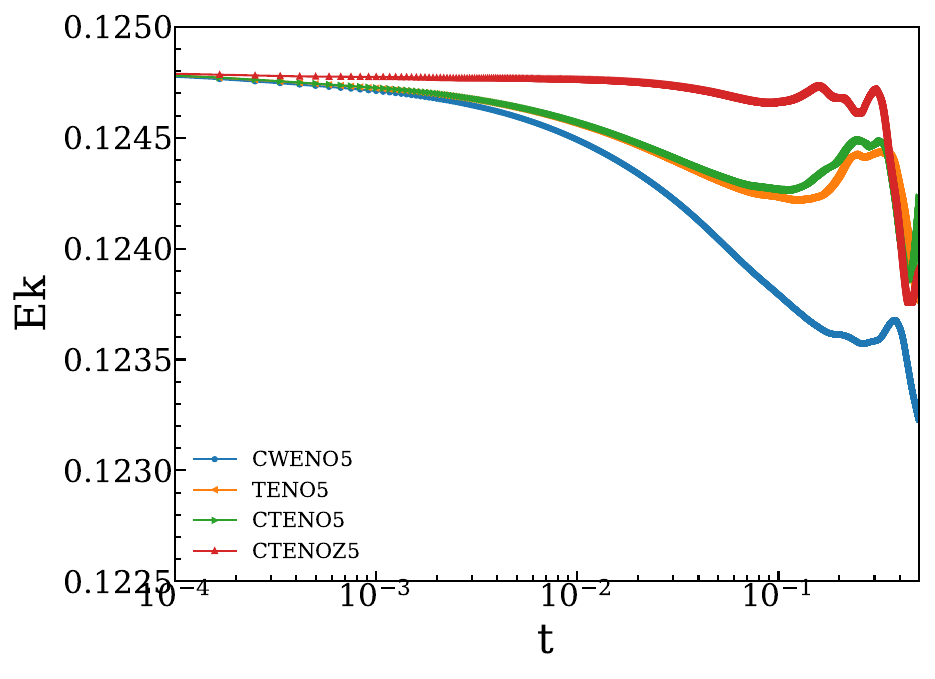}
   
	\caption{2D Kelvin-Helmholtz instability: Time histories of total kinetic energy \(E_{K}\) obtained with CWENO, TENO, CTENO,and CTENOZ schemes with 5th orders.}
	\label{figenergy}
\end{figure*}

When examining the development of overall kinetic energy \(E_{K}\) depicted in \autoref{figenergy}, it becomes apparent that CWENO exhibits the highest level of dissipation, whereas CTENOZ demonstrates the least dissipation. Due to the unstable development of the Kelvin-Helmholtz instability in the later stages, vortices and rolling occur, causing variations in the system’s total kinetic energy as vortices enter and leave the computational domain. The total kinetic energy experiences fluctuation of increase and decrease in the later stages. We focused on the early stages of vortex generation and variations in kinetic energy dissipation during the decline of total kinetic energy. The developed CTENO and CTENOZ schemes exhibited lower dissipation in comparison. When it comes to the conservation of kinetic energy, CTENO and CTENOZ outperform CWENO and TENO schemes. Therefore, $\lambda_{1} ^{'}=10^{4}$ is applied in the remained test problem for achieving 5th order of accuracy for CTENOZ.

\subsection{Interaction of a Shock Wave with a Wedge in 2D}
\label{}

\begin{figure*}[h]
	\centering
 
       \subfigure[]{
    \label{}
    \includegraphics[scale=0.295]{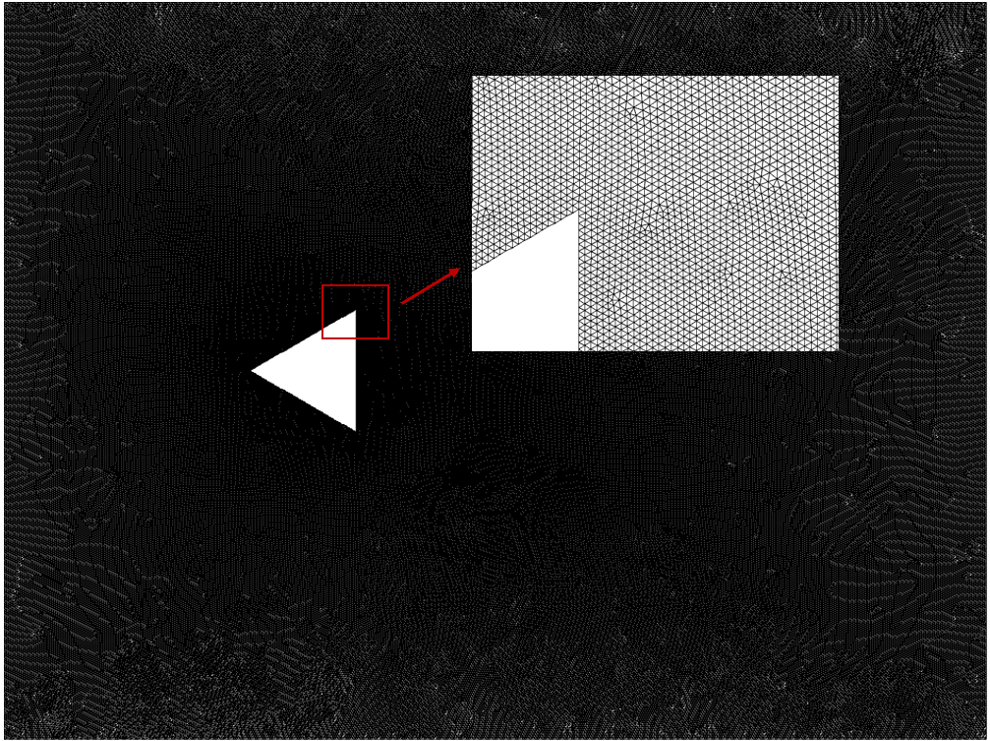}}
     \subfigure[]{
    	\label{}
    	\includegraphics[scale=0.33]{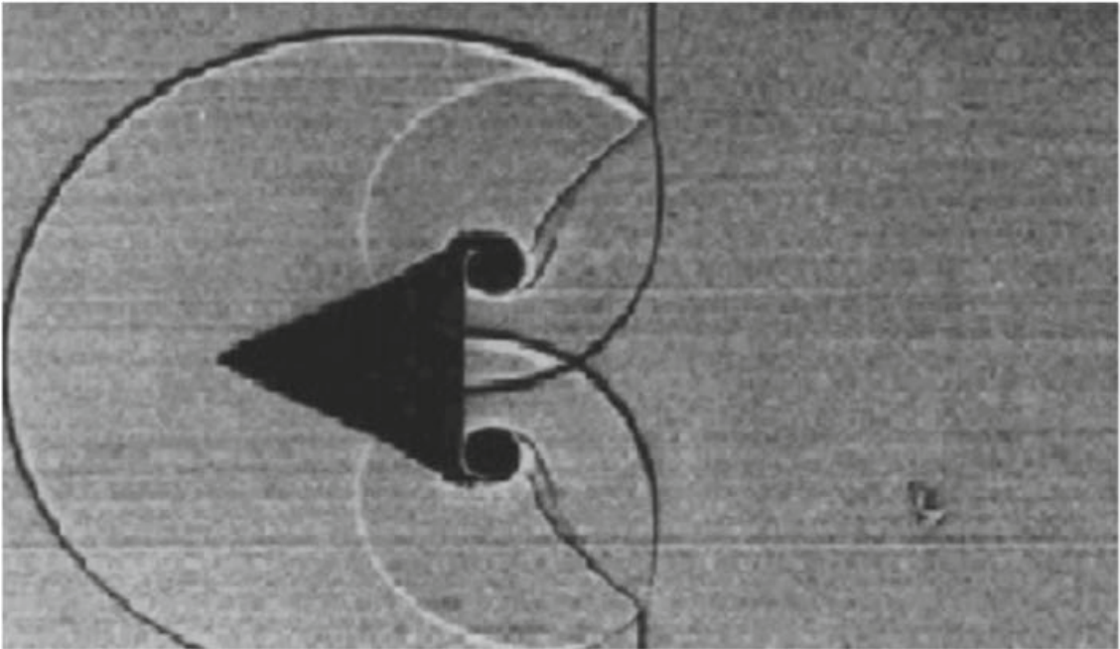}}
    
	\caption{Interaction of a shock wave with a wedge in 2D: (a) computational domain with 413791 uniform triangular elements. The region around wedge has being refined and (b) experiment at \(t = 2.1 \).}
	\label{fig10}
\end{figure*}

The Schearldin's probllem \cite{45}, which deals with interaction between shock wave and 2D wedge, is considered in this subsection. The simulation domain we are working with is \([-2,6] \times [-3,3]\), with the wedge tip positioned at \([0,0]\). The length and height of wedge has one unit, with symmetrical conditions imposed on top and bottom regions. The inflow and outflow conditions are assigned to the respective left and right sides. To enhance precision of capturing interaction, unstructured mesh has been refined near the wedge. \autoref{fig10} visualizes a mesh comprising 413791 uniformly distributed triangular elements in this test and the experimental results at \(t = 2.1\). The initial condition is provided according to \cite{46}.
\begin {equation}
(\rho ,u,v,p)=\begin{cases}
  (2.122,0.442,0,1.805), \ \text{ if } x< -1 \\
  (1.4,0,0,1). \ \text{ otherwise } 
\end{cases}
 \end {equation}

\begin{figure*}[h]
	\centering
 
    \subfigure[t=2.1]{
    \label{}
    \includegraphics[scale=0.2]{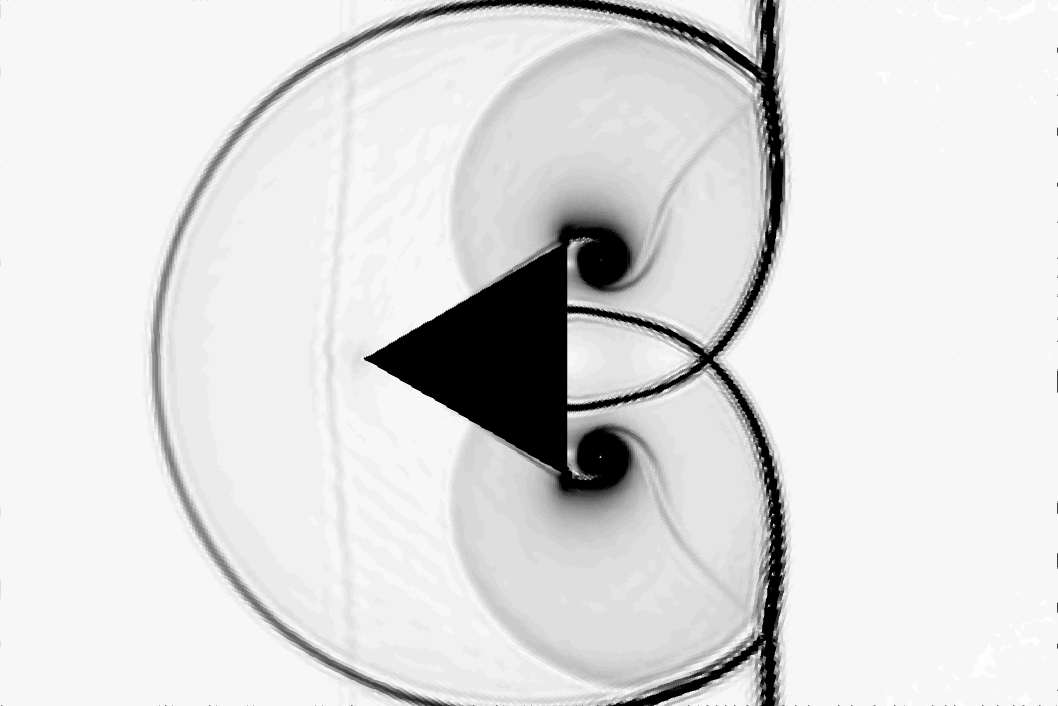}}
     \subfigure[t=4]{
     \label{}
    \includegraphics[scale=0.2]{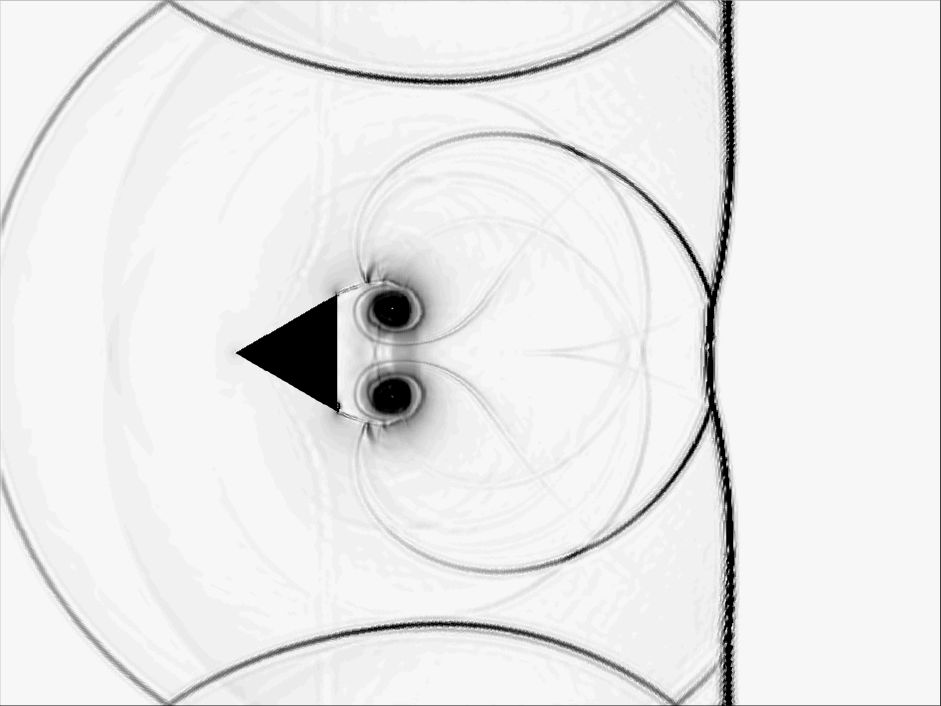}}
   
	\caption{Interaction of a shock wave with a wedge in 2D: result from CTENOZ5 scheme at \(t = 2.1 \) and \(t = 4 \). The figures contain 40 density gradient magnitude contours ranging from 0 to 5.}
   
	\label{fig11}
\end{figure*}

The density field computed at \(t = 2.1\) using the 5th order CTENOZ schemes is presented in \autoref{fig11}, demonstrating well agreement with experimental measurements. The shock wave patterns and counter-rotating vortices observed downstream at \(t = 4\) is also displayed with the proposed methods, exhibiting shock capture ability in the density fields. These findings indicate that the considered schemes accurately captures all significant flow features in their exact positions, including bow shock, tip vortices, and interface between them.

\subsection{Helium Bubble Shock Wave}
\label{}

This model has been created to simulate interaction between bubble and low-intensity shockwave, which is widely utilized to assess numerous methods for modeling multicomponent flow. In this study, quasi-conservative five-equation model is utilized to characterize behavior of inviscid compressible multicomponent flows, which is proposed by Allaire et al. \cite{47}. The equations are closed by incorporating a stiffened gas equation of state. A bubble, consisting of a mixture of helium and air, with a diameter of \(D = 5 cm\), is positioned inside a shock tube filled with air. Bubble is affected by the shockwave traveling from right to left, causing contamination of air. The value of specific heats of air and helium are assigned to values of 1.4 and 1.66 respectively. The initial state is defined in reference \cite{48}.
\begin{equation}
\begin{array}{c}
(a_{1}\rho_{1},a_{2}\rho_{2} ,u,v,p,a_{1}) = 
\begin{cases}
  (0.158,0.016,0,0,101325,0.95),  \ \text{ for Bubble }  \\
  (0,1.658,-114.49,0,159060,0), \ \text{ for Post-shock } \\
  (0,1.204,0,0,101325,0), \ \text{for Pre-shock }
\end{cases}
\end{array}
\end{equation}

\begin{figure*}[h]
	\centering
    \includegraphics[scale=0.4]{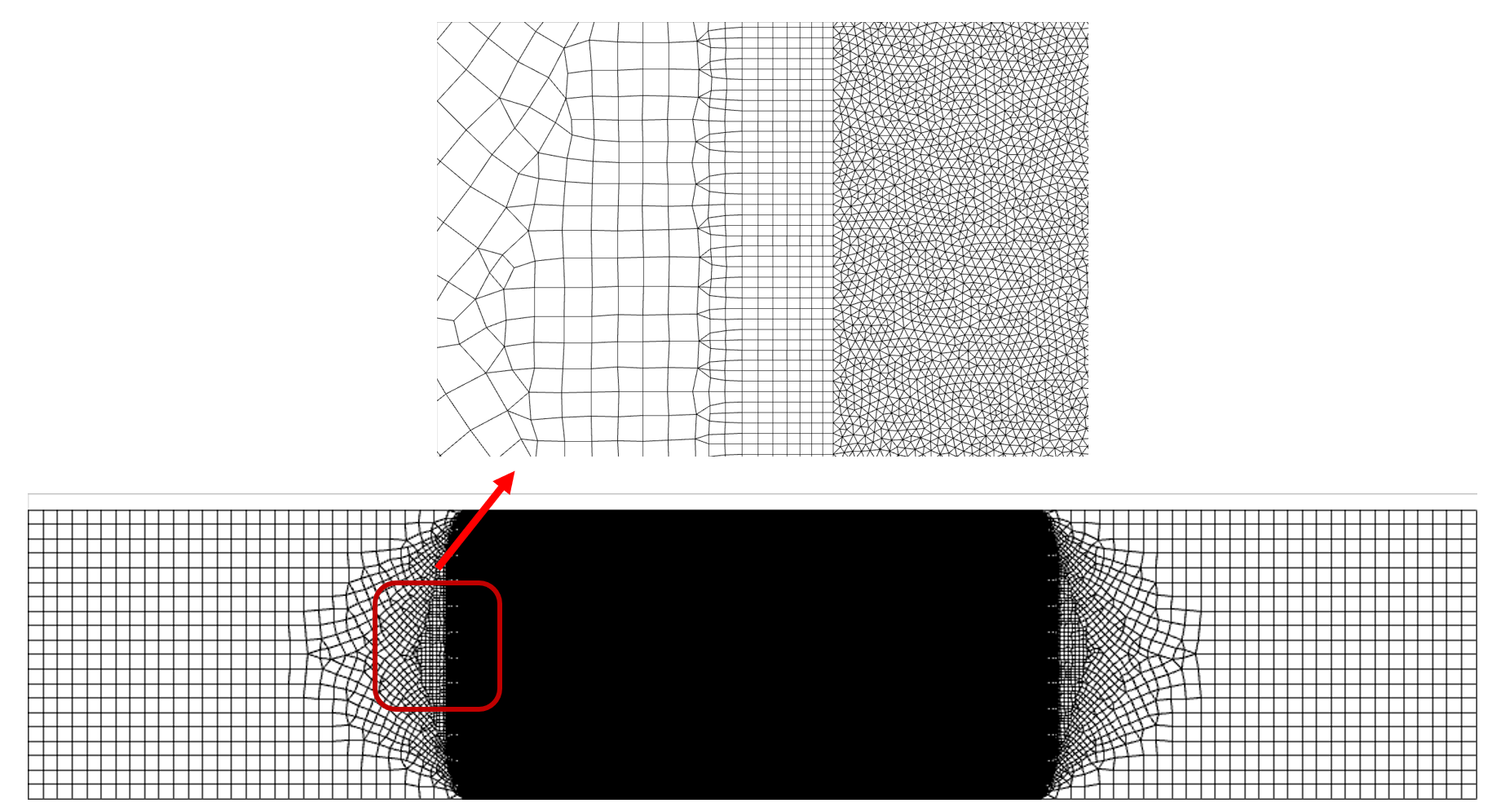}
	\caption{Hybrid grid for Helium Bubble Shock Wave.}
	\label{hybrid}
\end{figure*}

The computational domain utilized in this study is \([-0.25, 0.25] \times [0, 0.1]\). This domain is composed of three regions, pre-shock, post-shock, and bubble regions. We utilized a unstructured mesh with mixed-element consists of triangular and quadrilateral cells, as shown in \autoref{hybrid}. It is crucial that we refined shock-bubble interaction region with hybrid grid to improve the accuracy of our simulations. The total mesh consists of 256252 elements. The slip-wall conditions were imposed at the domain's top and bottom edges. Inflow and outflow conditions were utilized at right and left edges, respectively. We utilized CTENOZ scheme with a 5th-order accuracy, and the simulations were conducted until \(t=983\mu s\).

\begin{figure*}[h]
	\centering
 
    \subfigure[CTENOZ5   \(t=101\mu s\)]{
    \label{}
    \includegraphics[scale=0.135]{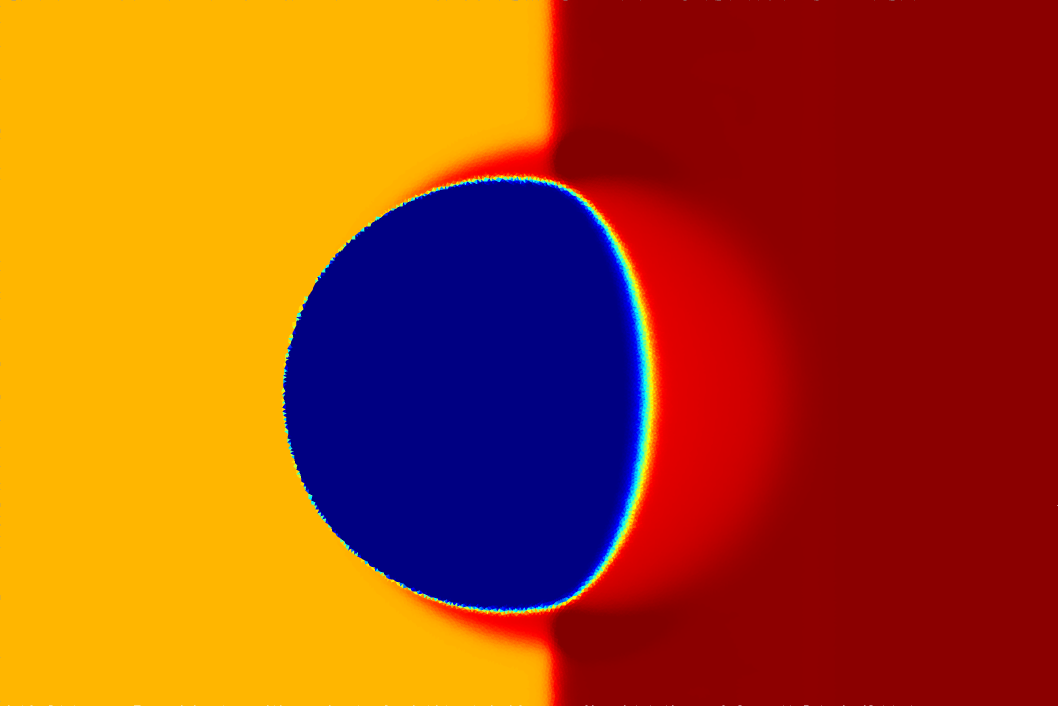}}
    \subfigure[CTENOZ5   \(t=400\mu s\)]{
    \label{}
    \includegraphics[scale=0.135]{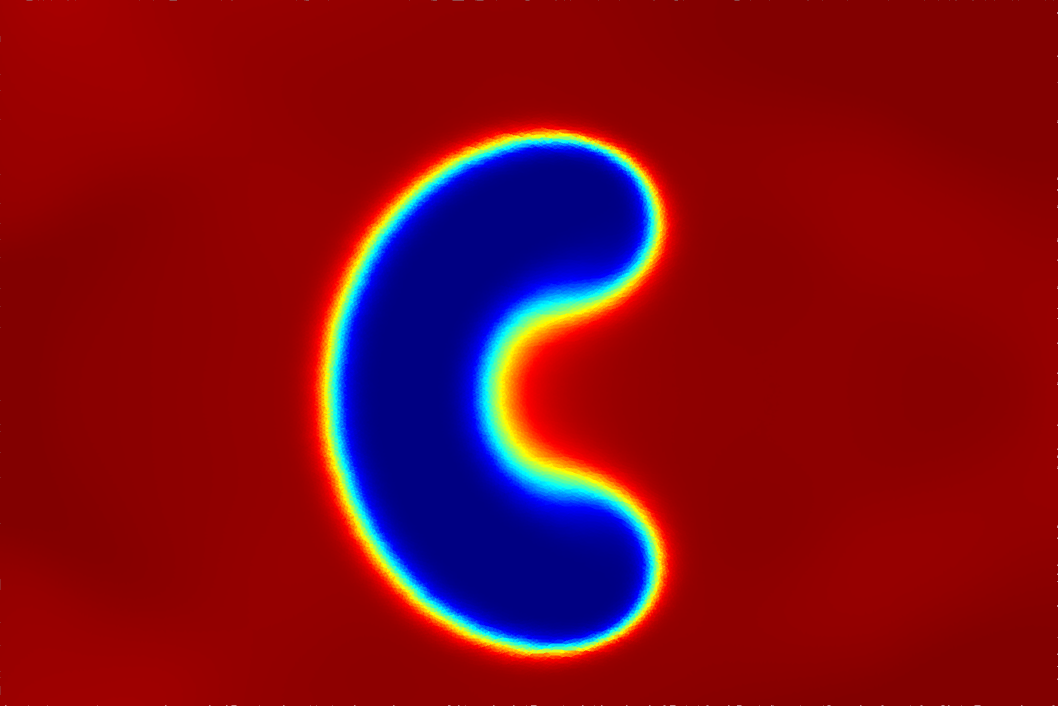}}
    \subfigure[CTENOZ5   \(t=983\mu s\)]{
    \label{}
    \includegraphics[scale=0.135]{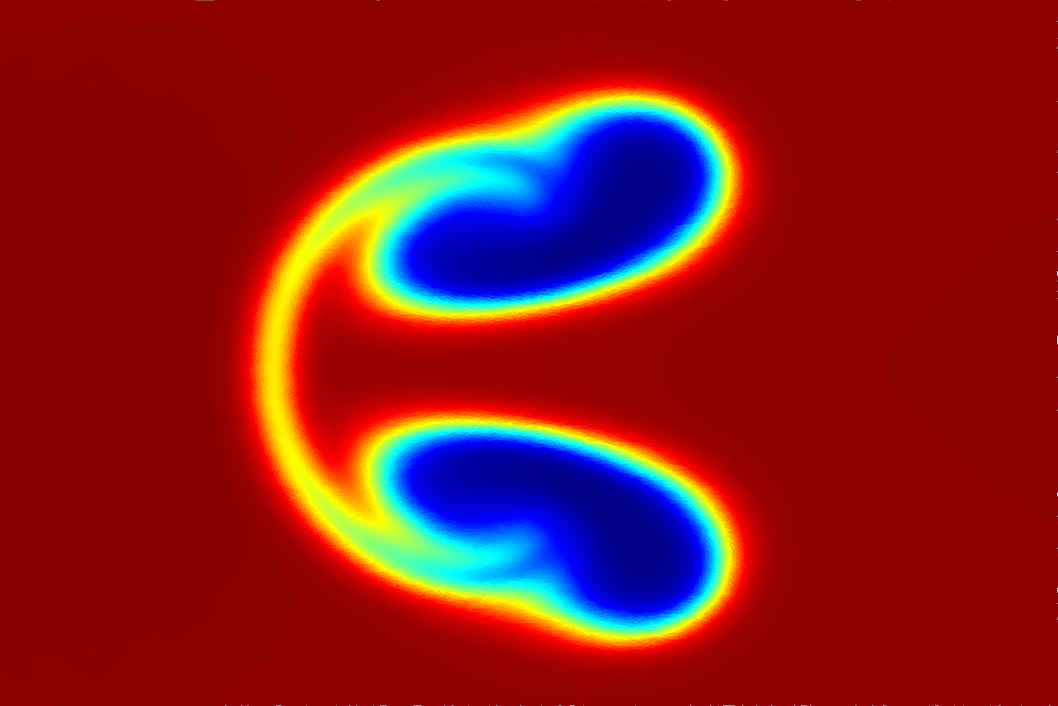}}
   
	\caption{Helium Bubble Shock Wave:results at various instants from the CTENOZ schemes with 5th-order. The figures contain 14 density contours ranging from 0.3 to 1.6.}
	\label{fig13}
\end{figure*}

According to the displayed results in \autoref{fig13}, it is clear that CTENOZ scheme successfully capture the interaction between the bubble and shockwave during the time evolution. The occurrence of late-stage phenomena, such as the creation of a vortex ring and a jet, has been recognized. These findings align with the qualitative outcomes observed in previous research studies \cite{48,49}. The instability at the helium bubble interface is amplified by the application of high-order schemes. The spatial-temporal positions of these interfaces reveals a close alignment between the predicted interface positions and the results of reference provided by Terashima and Tryggvason \cite{50} and Quirk and Karni \cite{51}.

\subsection{3D Explosion Problem}

We consider a 3D explosion problem to evaluate the high-order CTENOZ schemes on unstructured meshes for 3D tests. The computational domain consists of a sphere with a radius of one. The initial condition is given by

\begin{equation}
\begin{array}{c}
(\rho,u,v,w,p) = 
\begin{cases}
  (1,0,0,0,1),  \ \text{ if }\sqrt{x^{2}+y^{2}+z^{2}} \le 0.5  \\
  (0.125,0,0,0,0.1). \ \text{ otherwise } \\
\end{cases}
\end{array}
\end{equation}

\begin{figure*}[h]
	\centering
 
   \subfigure[]{
    \label{}
    \includegraphics[scale=0.25]{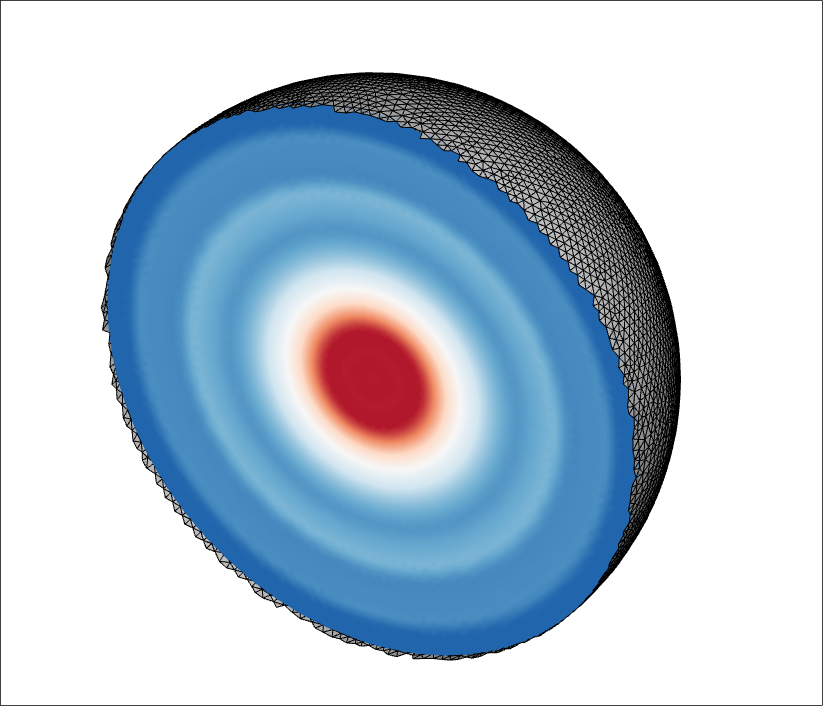}}
    \subfigure[]{
    \label{}
    \includegraphics[scale=0.25]{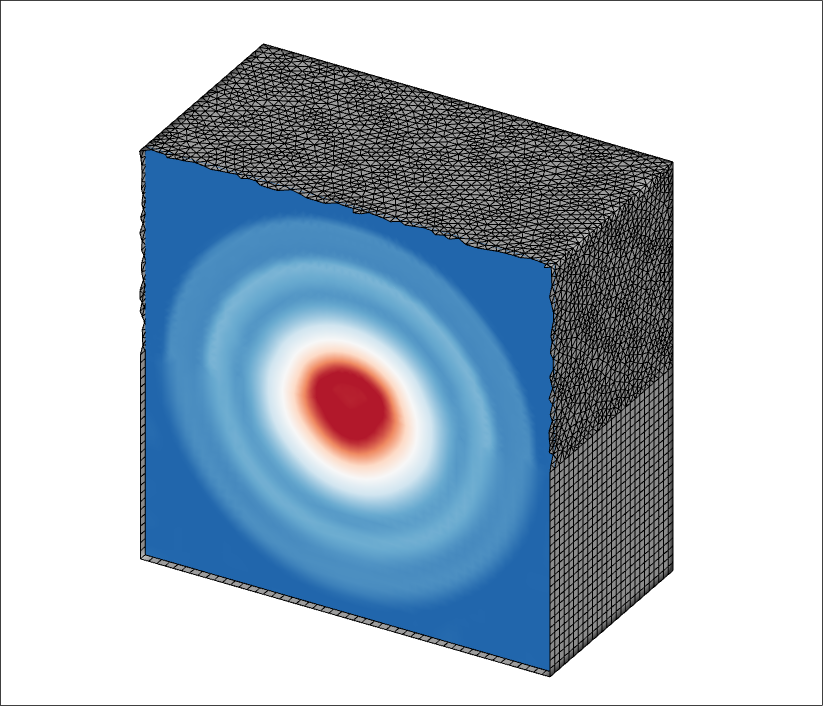}}
     \subfigure[]{
    \label{}
    \includegraphics[scale=0.45]{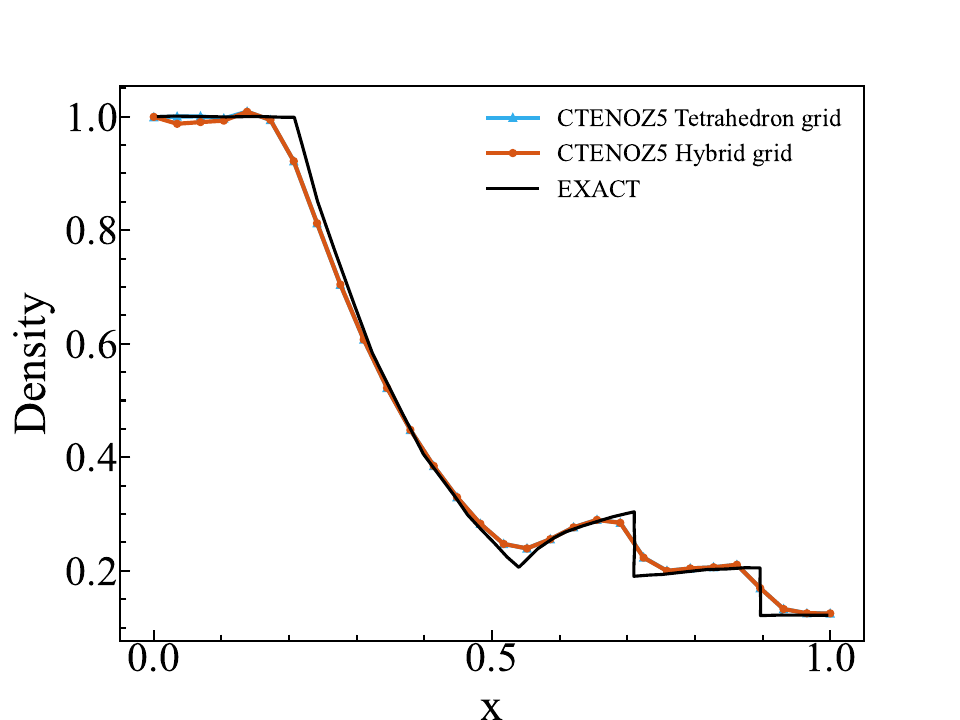}}
    \subfigure[]{
    \label{}
    \includegraphics[scale=0.45]{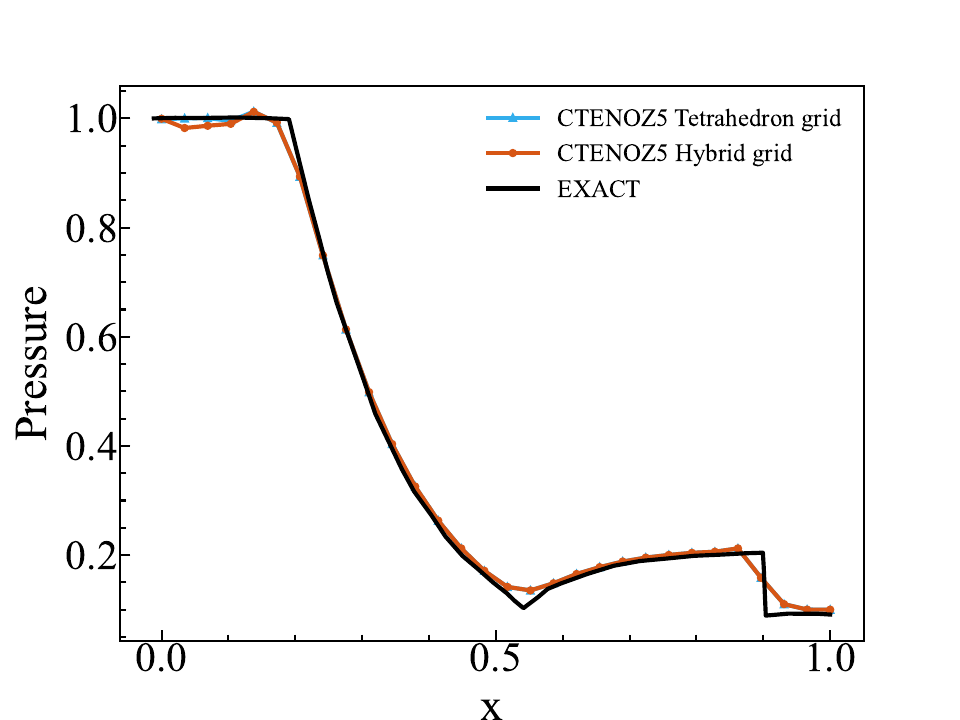}}
   
	\caption{3D Explosion Problem: (a)Unstructured tetrahedron elements with cross-sectional density distribution of simulation results, ranging from 0.15 to 1.0 with 18 levels. (b)Unstructured hybrid elements with cross-sectional density distribution of simulation results, ranging from 0.15 to 1.0 with 18 levels. (c)Density profile compared with exact results. (d)Pressure profile compared with exact results.}
	\label{fig14}
\end{figure*}

The final simulation time is $t = 0.25$. The unstructured mesh consists of 914546 uniformly tetrahedron cells were utilized for simulation. A hybrid mesh consists of 965027 elements is also tested in this problem. The grid of computational domain and cross-sectional density distribution of simulation results are depicted in \autoref{fig14}. Both density and pressure distribution from 5th order CTENOZ scheme agree well with the analytical exact solution in tetrahedron and hybrid mesh, which shows scalability for different meshes. 

\section{Conclusion}
\label{}

The TENO scheme has been extended to central targeted ENO (CTENO) family schemes for unstructured meshes, achieving high-order accuracy. The developed CTENO and CTENOZ schemes have been validated against a range of test cases and demonstrate robust high-order precision. These schemes feature a large central-biased stencil combined with multiple smaller directional stencils. Compared to WENO stencils of the same order, the CWENO, TENO, CTENO, and CTENOZ schemes have a more compact overall stencil width. Similar to the TENO scheme used for structured grids, these central TENO family schemes for unstructured grids maintain low numerical dissipation by carefully selecting, optimizing, or even disregarding stencils in the presence of discontinuities. To adapt to unstructured meshes, an ENO-like stencil selection strategy is employed, using a rigorous approach for strong scale separation. The CTENO and CTENOZ schemes ensure high-order accuracy by reconstructing from large central stencils in smooth regions while maintaining sharp shock-capturing capabilities by selecting candidate reconstructions from small directional stencils. The CTENO scheme further leverages optional polynomial and linear coefficients from CWENO schemes to maximize its potential for high-order accuracy. By incorporating smoothness indicators derived from polynomials of varying orders, the CTENOZ scheme achieves even lower dissipation compared to CTENO. These schemes are designed to achieve up to seventh-order accuracy, with their parameters explicitly defined to enhance performance.

Several benchmark simulations were carried out in order to assess the proposed schemes. The numerical findings reveal that these scheme exhibits great robustness in conducting compressible fluid simulations, with low dissipation and ability to accurately capture discontinuities. The numerical accuracy results shows that performance of CTENO scheme is influenced by the central stencil linear weight while CTENOZ scheme is less sensitive to it. Moreover, the CTENO and CTENOZ schemes exhibit reduced dissipation and enhanced accuracy compared to CWENO and TENO schemes of equivalent accuracy orders, while still maintaining their strong shock-capturing capability. The parallel scalability is tested and proves great scaling performance. The computational time of the proposed CTENO and CTENOZ can be roughly equivalent compared to CWENO and TENO schemes. Considering the impressive effectiveness demonstrated by the CTENO and CTENOZ framework and its potential for future expansion, future research will concentrate on implementing these methods in more complex flows, including chemical-reacting flows and external aerodynamics involving realistic geometries.

\section*{Acknowledgments}
Project supported by the National Natural Science Foundation of China under grant Nos. 11988102, 12432011, 12422208, 12372220 and 12421002.


\bibliographystyle{unsrt}
\bibliography{ref}




\end{document}